\documentclass[preprint, 3p, twocolumn]{elsarticle}

\usepackage[cmex10]{amsmath}	
\usepackage{amssymb}
\usepackage{siunitx}
\usepackage{bm}
\renewcommand*\descriptionlabel[1]{\hspace\leftmargin$#1$}

\usepackage[linesnumbered,ruled,vlined]{algorithm2e}
\SetKwInput{KwInput}{Input}              
\SetKwInput{KwOutput}{Output}            

\SetCommentSty{mycommfont}

\usepackage[capitalise]{cleveref}
\crefformat{equation}{(#2#1#3)}

\biboptions{compress}
\usepackage{url}

\usepackage{xcolor}

\usepackage{graphicx}
\usepackage[caption=false,font=footnotesize]{subfig}

\usepackage{tikz}
\usetikzlibrary{arrows}
\usetikzlibrary{shapes}

\newcommand{\mymk}[1]{
  \tikz[baseline=(char.base)]\node[anchor=south west, draw,rectangle, rounded corners, inner sep=0.1pt, minimum size=3.5mm,
    text height=2mm](char){\ensuremath{#1}} ;}

\newcommand*\circled[1]{\tikz[baseline=(char.base)]{
            \node[shape=circle,draw,inner sep=0.1pt] (char) {#1};}}
            
\newcommand*\rectangle[1]{\tikz[baseline=(char.base)]{
            \node[shape=rectangle,draw,inner sep=0.6pt] (char) {#1};}}
\makeatother

\graphicspath{{Figures/}}

\begin{document}

\begin{frontmatter}
		
	\title{Traffic-aware Gateway Placement and\\Queue Management in Flying Networks}
		
	\author{André Coelho\corref{cor1}}
	\ead{andre.f.coelho@inesctec.pt}
		
	\author{Rui Campos}
	\ead{rui.l.campos@inesctec.pt}
		
	\author{Manuel Ricardo}
	\ead{manuel.ricardo@inesctec.pt}
		
	\address{INESC TEC and Faculdade de Engenharia, Universidade do Porto, Portugal}
		
	\cortext[cor1]{Corresponding author at INESC TEC and Faculdade de Engenharia, Universidade do Porto, Campus da FEUP, Rua Dr. Roberto Frias, 4200-465 Porto, Portugal}

	%%%%%%%%%%%%%%%%%%%%%%%%%%%%%%%%%%%%%%%%%%%%%%%%%%%%%%%%%%%%%%%%%
	% ABSTRACT AND KEYWORDS
	%%%%%%%%%%%%%%%%%%%%%%%%%%%%%%%%%%%%%%%%%%%%%%%%%%%%%%%%%%%%%%%%%
	\begin{abstract}
				
		Unmanned Aerial Vehicles (UAVs) have emerged as adequate platforms to carry communications nodes, including Wi-Fi Access Points and cellular Base Stations. This has led to the concept of flying networks composed of UAVs as a flexible and agile solution to provide on-demand wireless connectivity anytime, anywhere. However, state of the art works have been focused on optimizing the placement of the access network providing connectivity to ground users, overlooking the backhaul network design. In order to improve the overall Quality of Service (QoS) offered to ground users, the placement of Flying Gateways (FGWs) and the size of the queues configured in the UAVs need to be carefully defined to meet strict performance requirements. The main contribution of this article is a traffic-aware gateway placement and queue management (GPQM) algorithm for flying networks. GPQM takes advantage of knowing in advance the positions of the UAVs and their traffic demand to determine the FGW position and the queue size of the UAVs, in order to maximize the aggregate throughput and provide stochastic delay guarantees. GPQM is evaluated by means of ns-3 simulations, considering a realistic wireless channel model. The results demonstrate significant gains in the QoS offered when GPQM is used.
				
	\end{abstract}
		
		
\end{frontmatter}

%%%%%%%%%%%%%%%%%%%%%%%%%%%%%%%%%%%%%%%%%%%%%
% INTRODUCTION
%%%%%%%%%%%%%%%%%%%%%%%%%%%%%%%%%%%%%%%%%%%%%
\section{Introduction} \label{sec:introduction}

\begin{figure}[!t]
	\centering
	\includegraphics[width=1\linewidth]{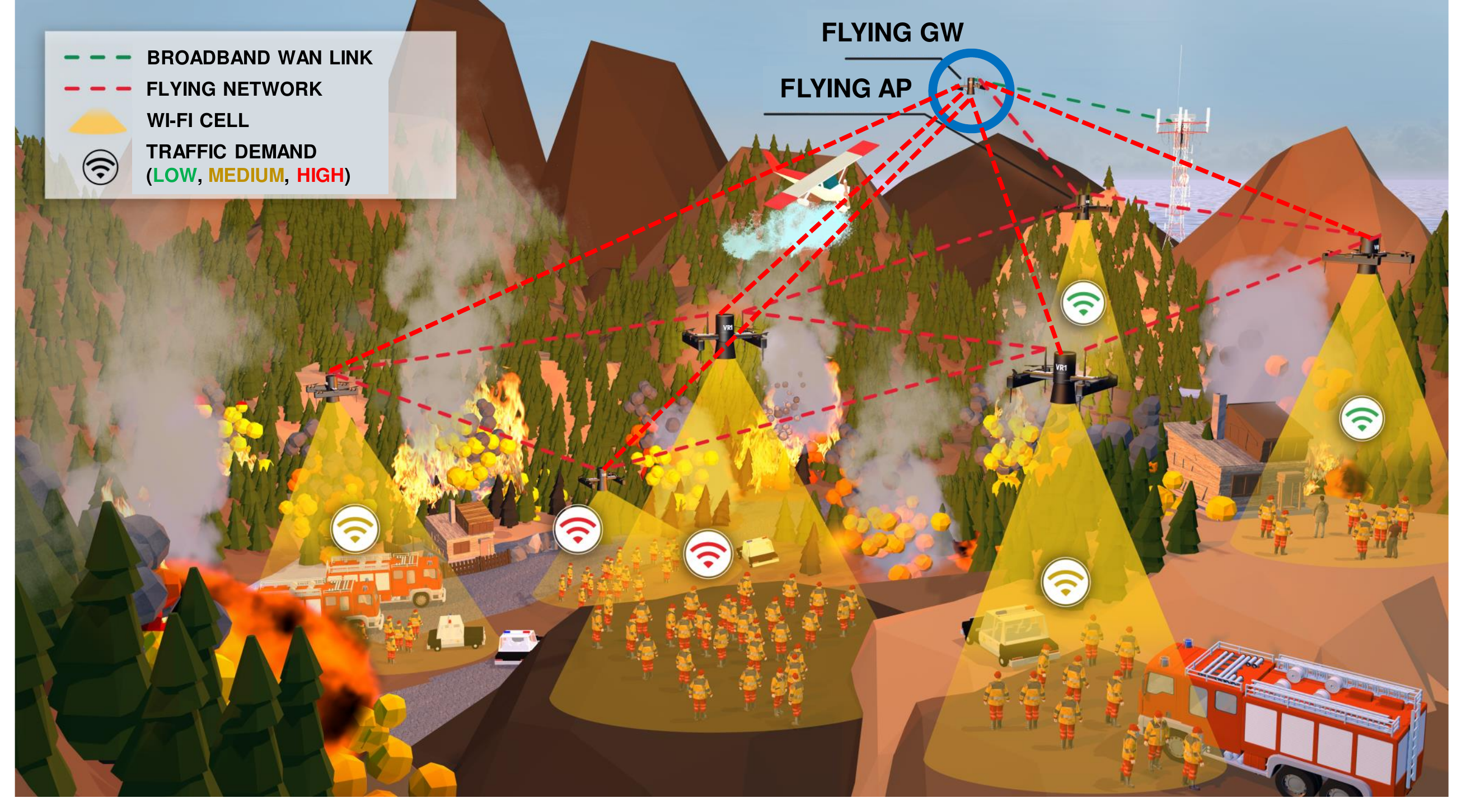}
	\caption{Flying network providing on-demand broadband wireless connectivity to first responders in a disaster management scenario.}
	\label{fig:flying-network-emergency-scenario}
\end{figure}

Flying networks, composed of Unmanned Aerial Vehicles (UAVs), have emerged as a flexible and agile solution to provide on-demand, temporary wireless connectivity when there is no network infrastructure available or there is a need to enhance the capacity of existing networks \cite{bor2016, zeng2016survey, Almeida2018, chakraborty2018, zhao2019} \cite{Pang2021}. Their ability to operate virtually anywhere and hover above the ground, their capability to move in 3D space, and their growing on-board payload capacity make UAVs adequate platforms to carry communications nodes, including Wi-Fi Access Points and cellular Base Stations \cite{Hayat2016}. The reference concept proposed in \cite{WISE:online} is depicted in \cref{fig:flying-network-emergency-scenario} for a disaster management scenario. It is based on 1) Flying Access Points (FAPs) that position autonomously according to the traffic demand of the ground users and establish an access network for coverage extension and capacity enhancement, and 2) a Flying Gateway (FGW) that forwards the traffic to/from the Internet, establishing a backhaul network with the FAPs. Yet, the ever increasing usage of a myriad of communications services and online applications, such as ultra-high definition video streaming, augmented and virtual reality, holographic-type communications, and remote vehicle piloting, poses orthogonal performance requirements to the access and backhaul networks, including high throughput and delay guarantees \cite{Simsek2016, alwahab2018, Clemm2020}. These requirements are envisioned in the emerging 5G+ networks, which aim at providing differentiated services to multiple verticals, including massive Machine-Type Communications (mMTC)~\cite{Shen2020, Sabuj2022}, massive Internet of Things (mIoT) communications~\cite{Liu2019, Liu2020, Liu2021}, Ultra-Reliable and Low-Latency Communications (URLLC)~\cite{Han2021, Cai2022}, and enhanced Mobile Broadband (eMBB)~\cite{Xi2021, Wu2022}, while ensuring target performance guarantees on top of a shared physical wireless network infrastructure.

When it comes to the FAP placement problem, existing solutions are focused on enhancing the radio coverage and the number of ground users served, as well as improving the Quality of Service (QoS) offered by the access network. Even though the ground users are directly affected by the access network, the backhaul network -- including the UAVs acting as network relays and gateways to/from the Internet -- needs to be carefully designed. Still, this aspect has been overlooked in the state of the art, where an unconstrained backhaul network is commonly considered \cite{Almeida2018}. In order to address this challenge, in \cite{Coelho2019GWP} we have proposed a centralized traffic-aware gateway placement (GWP) algorithm for flying networks with controlled topology, which enables backhaul communications paths with high enough capacity to accommodate the traffic offered by the access network. 

In addition to the ability to accommodate the offered traffic, flying networks providing wireless connectivity are expected to meet strict low-latency requirements~\cite{Masaracchia2021}. To overcome this challenge, the flexibility provided by UAVs to be placed so that short-distance Line of Sight wireless links are ensured is a step forward~\cite{She2018}. Yet, unprecedented challenges from the network configuration point of view are also arising. In particular, the size of the queues configured in the communications nodes, which are used for accommodating transient traffic bursts offered to the network, needs to be defined.

Wireless networks in general and flying networks in particular, due to their dynamic nature, bring up significant challenges, including: 1) shared wireless channels with capacity constrained by the number of neighboring nodes; and 2) frequent changes in the Signal to Noise Ratio (SNR) of the wireless links, which induce time-varying network capacity. For these reasons, queues with static size and the state of the art Active Queue Management (AQM) algorithms~\cite{floyd1993, feng1999, hollot2001, nichols2012, pan2013}, which were designed for wired networks, are typically not adequate for flying networks. 
This challenge was overcome in \cite{Coelho2020}, where we have proposed a Proactive Queue Management (PQM) algorithm able to provide stochastic delay guarantees. However, PQM is limited to the proactive configuration of the queue size. An algorithm able to ensure wireless links that meet the traffic demand of the FAPs, while benefiting from the network capacity information for computing the queue size so that strict delay requirements are met, is worthy of being considered.

This article proposes a traffic-aware gateway placement and queue management algorithm, called GPQM. Taking advantage of the information on the future positions of the FAPs, defined according to the traffic demand of the ground users by a state of the art FAP placement algorithm running in a central station, the GPQM algorithm is able to 1) determine the FGW position, in order to accommodate the offered traffic, and 2) compute the queue size for each FAP, in order to provide stochastic delay guarantees. 

The main contributions of this article are two-fold:
\begin{enumerate}
	\item \textbf{A traffic-aware gateway placement and queue management algorithm for flying networks, called GPQM}. Building upon the GWP~\cite{Coelho2019GWP} and PQM~\cite{Coelho2020} algorithms, GPQM allows to jointly define an adequate placement for the FGW and a suitable queue size for the FAPs, which are aspects addressed by GWP and PQM independently, without taking advantage of the gains allowed by each other. GPQM allows to minimize the packet service time and the traffic load in each FAP, while improving the flying network performance.
	\item \textbf{Evaluation of the GPQM algorithm under realistic channel conditions}. GPQM is evaluated by means of ns-3 simulations \cite{ns-3}, considering different traffic generation models and a realistic wireless channel model, which allows to achieve more accurate and realistic performance results than using pure theoretical models. 
\end{enumerate}

The rest of this article is organized as follows.
\cref{sec:RelatedWork} discusses the related work.
\cref{sec:SystemModel} defines the system model and formulates the problem.
\cref{sec:GPQM} presents the GPQM algorithm.
\cref{sec:PerformanceEvaluation} describes the evaluation of the GPQM algorithm, including the methodology, the simulation setup, and the obtained results.
\cref{sec:Discussion} discusses the pros and cons of the GPQM algorithm.
Finally, \cref{sec:Conclusions} points out the main conclusions and directions for future work.

%%%%%%%%%%%%%%%%%%%%%%%%%%%%%%%%%%%%%%%%%%%%%
% RELATED WORK
%%%%%%%%%%%%%%%%%%%%%%%%%%%%%%%%%%%%%%%%%%%%%
\section{Related Work \label{sec:RelatedWork}}

Flying networks have been widely proposed in the literature as a suitable solution for providing communications in extreme scenarios. Associated with this, unprecedented challenges regarding UAV placement and flying network configuration, including queue management, have emerged. In this section, we review the related works on placement and queue management in wireless networks, with special focus on flying networks composed of UAVs.

Different solutions on gateway placement in wireless networks in general have been proposed \cite{maolin2009gateways, seyedzadegan2013zero, targon2010joint, aoun2006gateway, jahanshahi2019gateway, oueis2019}. Nevertheless, most of them aim at minimizing the number of gateways in scenarios with multiple gateways and do not take advantage of the controlled mobility over the communications nodes. A reference work is presented in \cite{muthaiah2008}, where it is shown that the gateway placement and the transmission power affect significantly the network throughput. When it comes to flying networks, the placement of UAVs acting as network relays and its impact on the achieved throughput has been studied in \cite{larsen2017, Chen2018}; yet, only a pair of ground nodes has been considered. A similar work employing a UAV in movement is presented in \cite{ono2016}. In \cite{zhong2019, zhong2020}, a model-free algorithm to determine the optimal position of UAVs acting as network relays is presented. Since it relies on online learning and interaction with the environment by means of real-time measurements, its main drawback is the long convergence time required to achieve the optimal solution. Overall, different placement algorithms for UAVs carrying communications nodes have been proposed in the literature \cite{cicek2019}. However, they are focused on UAVs acting as Access Points and Base Stations, aiming at maximizing the radio coverage and the number of ground users served \cite{mozaffari2015, kalantari2017, arribas2019, Sabzehali2021} or the QoS offered to the ground users \cite{Almeida2019, zeng2016, alzenad2018}. A reference work is presented in \cite{Almeida2018}, where an algorithm in charge of dynamically determining the coordinates to place the FAPs according to the traffic offered by the ground users is proposed, aiming at improving the aggregate throughput without compromising the coverage of the venue. Besides the placement of the UAVs forming the access network, an additional challenge in flying networks is the design of the backhaul network, which should ensure wireless links with high enough capacity to transport the traffic between the FAPs and the UAVs acting as network relays and gateways to/from the Internet.

The queues' size has also a significant impact on the QoS offered by a network. In order to overcome the challenge, different AQM algorithms have been proposed in the literature. They aim at dropping packets based on different approaches, including queue size, packet size, queuing delay, link utilization, and traffic classification. Some examples are Drop-Tail, Random Early Detection (RED) \cite{floyd1993}, BLUE \cite{feng1999}, Proportional Integral (PI) \cite{hollot2001}, Controlled Delay (CoDel) \cite{nichols2012}, and Proportional Integral controller Enhanced (PIE) \cite{pan2013}. In Drop-Tail, the size of the queue is set to a static value; the packets are accepted until this value is reached. The main disadvantage of Drop-Tail lies in the fact that packets can be held in the queue for long time if the size of the queue is not carefully defined. RED drops packets based on queue size. It employs minimum and maximum threshold values for the queue size. When the average queue size is between the defined threshold values, the packets are marked with a dropping probability. In turn, when the number of packets in the queue is higher than the maximum threshold value, all packets are dropped. However, RED requires fine-tuning several parameters, including the minimum and maximum threshold values for the queue size and the probability for marking and dropping packets when congestion occurs, which need to be adjusted according to the networking scenario \cite{Feng1999RED}. This makes RED not suitable for wireless networks with frequent topology changes, which lead to links with time-varying capacity. BLUE considers previous link under-utilization and packet loss events to manage network congestion. It randomly drops incoming packets based on a drop probability, which is incremented when a packet is lost and decremented when the link is idle. Yet, BLUE was designed to tackle non-responsive traffic only. PI employs a proportional integral to control the queue size. Initially, the queue size is defined to a desirable value, and thereafter PI controls the queue size by updating the packet drop probability. However, PI reacts slowly to changes; this makes it not suitable for bursty traffic and wireless networks, which are dynamic in nature. Contrary to the previous AQM algorithms, CoDel, which has been recently proposed, relies on the sojourn time -- the time that packets are held in the queue. CoDel starts dropping packets at the head of the queue when the sojourn time is higher than a predefined value. However, CoDel requires per-packet time-stamping. Similarly to CoDel, PIE aims at controlling the average queuing delay. For that purpose, it randomly drops incoming packets at the onset of congestion, which is inferred based on the packet queuing delay. Still, the state of the art works have been focused on addressing the queue management in wired networks only~\cite{karakus2017}.

Wireless networks introduce additional challenges that must be taken into account, including time-varying wireless channel capacity and shared wireless channels \cite{showail2016}. These challenges are exacerbated in flying networks, which induce frequent topology changes. With the growing usage of communications services and online applications with strict performance requirements regarding throughput and delay, this research topic is emerging again \cite{alwahab2018, jain2018, bouacida2018}. In \cite{Li2010buffer}, the authors have studied the impact of the queue size in single-hop IEEE 802.11 networks. The performance of three different algorithms has been evaluated: emulated Bandwidth-Delay Product, Adaptive Limit Tuning (ALT), and A*. The authors have concluded that the usage of static size queues in IEEE 802.11 networks leads to underutilized communications nodes and unnecessary high delays. In \cite{showail2014}, an aggregation-aware queue management scheme for wireless networks (WQM) is presented. WQM estimates the time required to drain the queues based on the transmission rate, in order to adjust their size and meet the desired QoS. Using a Linux implementation, the authors have shown that WQM is able to reduce the delay 8 times when compared to the default Linux queues, at the cost of up to 15\% reduction in goodput. A queue management algorithm is presented in \cite{kulkarni2006}. It employs the Recursive Least Square (RLS) algorithm to predict the average queue size over time, which includes a set of dynamic parameters that are estimated in real-time, considering the queue occupancy observed in the previous time instants. This can lead to unstable behavior and compromise performance in wireless links with time-varying capacity. In~\cite{Zhang2021}, a latency-aware UAV placement algorithm has been proposed. Considering a network composed of a terrestrial Base Station and a UAV also acting as a Base Station, connected to each other by means of a Free-Space Optics-based backhaul link, the proposed algorithm aims at determining the UAV 3D position, user association, and bandwidth allocation between the UAV and terrestrial Base Station. The objective is to minimize the ground users' average latency and meet their QoS requirements; however, the queue management problem is not addressed.
A joint UAV placement and queue scheduling algorithm for admission and congestion control has been proposed in \cite{sharma2019}. Yet, it is focused on multiple UAVs that act as network relays for ground nodes, and the queue scheduling lies on assigning different priorities to the network packets. In \cite{fadlullah2016}, a dynamic trajectory control algorithm for UAVs providing wireless connectivity to ground nodes has been proposed. It defines the trajectories of the UAVs so that the queue occupancy of each UAV is lower than a given threshold, in order to increase the packet delivery ratio for the wireless links established between the UAVs. However, the trajectories of the UAVs are defined to ensure radio coverage for the ground nodes and the QoS requirements are not taken into account. In addition, since the trajectories of the UAVs are readjusted only when link congestion occurs, it constitutes a reactive approach. In \cite{Bai2020}, an algorithm for placing a UAV acting as a base station has been proposed. It aims at minimizing the average queuing delay with energy consumption and UAV speed constraints; however, the traffic demand and the delay requirements of the ground nodes are not considered. Overall, as stated in \cite{showail2016}, designing a single optimal queue management algorithm that suits the variable characteristics of wireless networks for different congestion scenarios is not straightforward, especially due to the need of fine-tuning a wide set of parameters. In \cite{li2013}, the authors have concluded that estimating the mean packet delay using the M/M/1 queuing model is accurate for flying networks under non-saturated traffic conditions and considering an independent packet error rate. This is an important conclusion that supports part of the work presented in our article.

%%%%%%%%%%%%%%%%%%%%%%%%%%%%%%%%%%%%%%%%%%%%%
% SYSTEM MODEL AND PROBLEM FORMULATION
%%%%%%%%%%%%%%%%%%%%%%%%%%%%%%%%%%%%%%%%%%%%%
\section{System Model and Problem Formulation \label{sec:SystemModel}}
\begin{figure}[!t]
	\centering
	\includegraphics[width=1\linewidth]{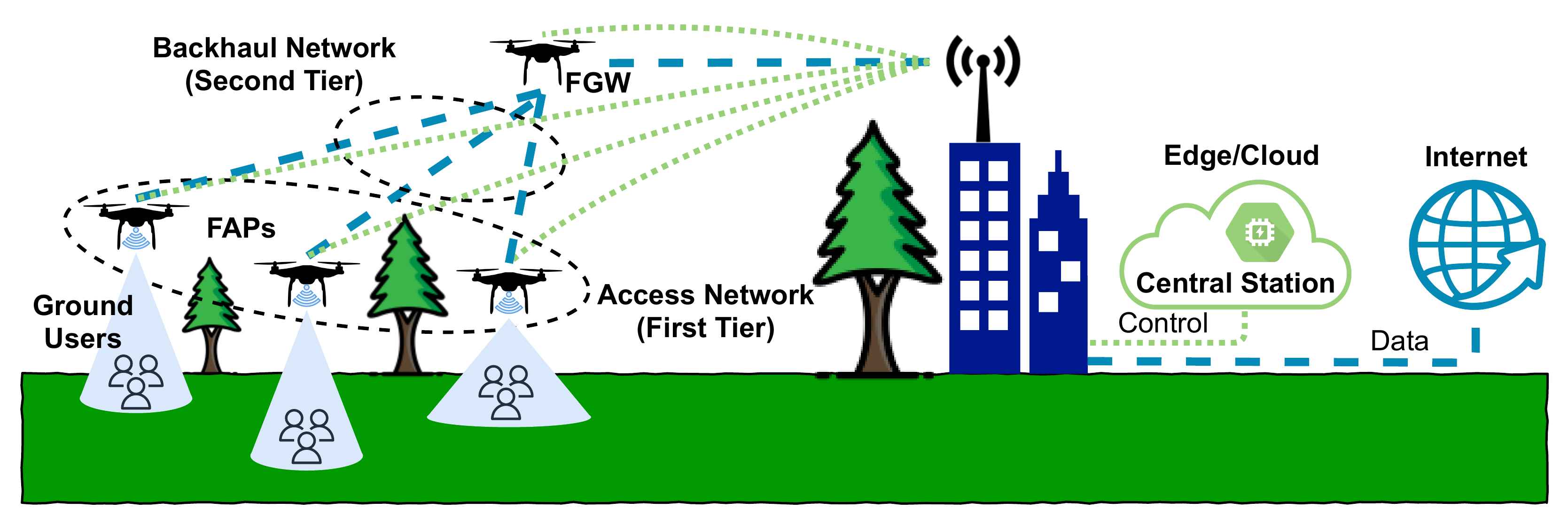}
	\caption{Flying network organized into a two-tier architecture. The FAPs provide wireless connectivity to the ground users and the FGW forwards the data traffic to/from the Internet (dashed blue line). The flying network is controlled by a Central Station deployed in the Edge or in the Cloud. The UAVs use an out-of-band long-range wireless channel (dotted green line) to exchange information with the Central Station.}
	\label{fig:flying-network-architecture}
\end{figure}

The flying network is assumed to be organized into a two-tier architecture, as depicted in \cref{fig:flying-network-architecture}. This architecture is especially targeted at taking advantage of short range high-directional wireless links, providing high bandwidth wireless channels and low inter-flow interference. Two types of UAVs are assumed to compose the flying network: 1) the FAPs, which provide wireless connectivity to the ground users, and 2) the FGW, which forwards the traffic to/from the Internet. The FAPs and the FGW establish a flying network able to 1) extend the radio coverage from a remote Access Point/Base Station and 2) provide on-demand wireless connectivity over a given area. The first tier consists of the access network, which is composed of FAPs enabling high-capacity small cells. The FAPs are dynamically placed and configured according to the traffic demand of the ground users by means of a state of the art FAP placement algorithm. For that purpose, the FAPs periodically take a snapshot of the first tier, including the amount of traffic that the ground users are offering to the network and send this information to a Central Station, where the FAP placement algorithm is running. The second tier is composed of the FGW that is placed above the FAPs. The FGW forwards the traffic to/from the Internet by means of a dedicated broadband wireless link with QoS ensured by the core network, typically managed by a telecommunications operator. The traffic demand of each FAP must be higher than or equal to the sum of the traffic offered by the ground users connected to it.

The Central Station can be deployed anywhere in the Edge or in the Cloud. The Central Station is responsible for controlling the flying network by periodically: 1) defining the positions of the FAPs according to the traffic offered by the ground users; 2) determining the updated positions of the FGW, in order to enable wireless links able to meet the traffic demand of the FAPs; and 3) calculating the queue size for each FAP, in order to provide stochastic delay guarantees. The Central Station is in charge of sending the updated positions to the FAPs and the FGW, and the queue size to each FAP, at periodic time instants. Based on the information received, the FAPs and the FGW position and configure themselves accordingly. An out-of-band long-range wireless channel (e.g., based on IEEE 802.11ah) is used for exchanging information between the Central Station and the UAVs. It enables an always-on control link, even when the flying network is being reconfigured, and avoids introducing overhead in the backhaul and access networks formed by the UAVs. The design of the out-of-band communications solution is beyond the scope of this article.

The main notation used to formulate the problem addressed by this article is presented in~\cref{tab:notation}. Let us assume that $\Delta t \gg \SI{1}{\second}$ corresponds to the update period defined by the FAP placement algorithm. At time $t_k = k \cdot \Delta t, k \in  \mathbb{N}_0$, the flying network is represented by a directed graph $G(t_k)=(V, E(t_k))$, where $V=\{0, ..., N-1\}$ is the set of UAVs $i$ placed at $P_i=(x_i, y_i, z_i)$ inside a cuboid $x^{MAX}$ long, $y^{MAX}$ wide, and $z^{MAX}$ high. $E(t_k) \subseteq  V \times V $ is the set of directional wireless links available between $\text{{UAV}}_i$ and $\text{{UAV}}_j$ at time $t_k$, with $(i,j)\in E(t_k)$. The wireless channel between each pair of UAVs is modeled by the Friis propagation loss model~\cite{friis1946note}, which is justified by the strong Line of Sight component that dominates the wireless links established between UAVs~\cite{Almeida2020}. $C_{j,i}(t_k)$ defines the maximum capacity, in bit/s, of the wireless link available between $\text{{UAV}}_i$ and $\text{{UAV}}_j$ at time $t_k$, considering a constant channel bandwidth $B$ in \SI{}{\hertz}. We assume that the maximum link capacity is equal to the data rate associated to the Modulation and Coding Scheme (MCS) index $\text{{MCS}}_i$ selected by each $\text{{UAV}}_i$ over the number of UAVs that use the same wireless channel. $\text{{MCS}}_i$ requires a minimum $\text{SNR}_i$, considering a constant noise power $P_N$.

\begin{table}
	\centering
	\caption{Main notation used to formulate the problem addressed by this article.}
	\label{tab:notation}
	\begin{tabular}{lp{6cm}lp{1cm}}
		\hline
		$B$              & Constant channel bandwidth, in \SI{}{\hertz}                                                                                                               
		\\
		$c$              & Speed of light in vacuum, in \SI{}{\meter/\second}                                                                                                         
		\\
		$C_{j,i}(t_k)$   & Maximum capacity, in \SI{}{bit/s}, of the wireless link available between $\text{{UAV}}_i$ and $\text{{UAV}}_j$, at time $t_k$                             
		\\
		$C^{MAX}$        & Maximum capacity, in \SI{}{bit/s}, of the wireless communications channel                                                                                  
		\\
		$D_i(t_k)$       & Average delay, in seconds, of the packets belonging to flow $F_{0,i}$, at time $t_k$                                                                       
		\\
		$d_{max_i}$      & Maximum distance, in meters, between $\text{{UAV}}_i$ and $\text{{UAV}}_0$ to ensure the selection of $\text{{MCS}}_i$                                     
		\\
		$E_T$            & Set of wireless links between $\text{{UAV}}_i$ and $\text{{UAV}}_0$                                                                                        
		\\
		$E(t_k)$         & Set of directional wireless links between $\text{{UAV}}_i$ and $\text{{UAV}}_j$, at time $t_k$                                                             
		\\
		$f$              & Carrier frequency, in \SI{}{\hertz}                                                                                                                        
		\\
		$F_{0,i}$        & Traffic flow transmitted by $\text{{UAV}}_i$ towards $\text{{UAV}}_0$                                                                                      
		\\
		$G(t_k)$         & Directed graph representing the flying network, at time $t_k$                                                                                              
		\\      
		$H$              & Maximum average delay allowed, in seconds                                                                                                                  
		\\
		$L$              & Reference fair share, in~\SI{}{bit/s}, of the wireless channel's capacity                                                                                  
		\\
		$M$              & Distance, in meters, to be ensured between the position computed for $\text{{UAV}}_0$ and the position of any other $\text{{UAV}}_i$ in the flying network 
		\\
		$\text{{MCS}}_i$ & Modulation and Coding Scheme (MCS) index selected by $\text{{UAV}}_i$                                                                                      
		\\    
		$P_i$            & Position of $\text{{UAV}}_i$                                                                                                                               
		\\
		$P_N$            & Constant noise power, in \SI{}{dBm}                                                                                                                        
		\\
		$P_T$            & Transmission power, in \SI{}{dBm}, of the UAVs                                                                                                             
		\\
		$Q_i$            & Queue size, in packets, of $\text{{UAV}}_i$                                                                                                                
		\\
		$R_i(t_k)$       & Bitrate, in~\SI{}{bit/s}, of the traffic flow $F_{0,i}$ measured in $\text{{UAV}}_0$, at time $t_k$                                                        
		\\  
		$S$              & Packet size, in bytes                                                                                                                                      
		\\
		$S_G$            & Gateway placement subspace                                                                                                                                 
		\\
		$\text{SNR}_i$   & Minimum Signal to Noise Ratio (SNR), in \SI{}{\deci\bel}, required for selecting $\text{{MCS}}_i$                                                          
		\\
		$T_i(t_k)$       & Bitrate, in~\SI{}{bit/s}, of the traffic flow transmitted by $\text{{UAV}}_i$ towards $\text{{UAV}}_0$, at time $t_k$                                      
		\\
		$T(V,E_T)$       & Active topology of the flying network                                                                                                                      
		\\
		$\text{{UAV}}_i$ & Unmanned Aerial Vehicle $i$                                                                                                                                
		\\
		$V$              & Set of UAVs composing the flying network                                                                                                                   
		\\
		$x_i,y_i,z_i$    & 3D Cartesian coordinates of $\text{{UAV}}_i$                                                                                                               
		\\
		\hline
	\end{tabular}
\end{table}

$\text{{UAV}}_i$, $i\in \{1, ..., N-1\}$, which is placed one hop away from $\text{{UAV}}_0$, acts as a FAP and transmits, during time interval $t_k$, a traffic flow with bitrate $T_i(t_k)$~\SI{}{bit/s} and constant packet size $S$, in bytes, towards $\text{{UAV}}_0$, which performs the role of FGW. The active topology of the flying network is defined by a tree $T(V, E_T)$ that is a subgraph of $G$, where $E_T \subseteq E$ is the set of direct wireless links between $\text{{UAV}}_i$ and $\text{{UAV}}_0$, which is the root of the tree. The flow $F_{0,i}$, with bitrate $T_i(t_k)$~\SI{}{bit/s}, demands a minimum capacity $C_{0,i}(t_k)$, in bit/s, for the wireless link available from $\text{{UAV}}_i$ to $\text{{UAV}}_0$, at time $t_k$. $F_{0,i}$ is received at $\text{{UAV}}_0$ from $\text{{UAV}}_i$ with bitrate $R_i(t_k)$ bit/s and average delay $D_i (t_k)$ seconds, at time $t_k$. Herein, we consider average delay values for illustrative purposes, but a Service Level Agreement established with a Mobile Network Operator can also refer minimum values (e.g., lowest packet delay among all packet delay values) or median values (e.g., 50\textsuperscript{th} percentile of the frequency distribution of packet delays)~\cite{Janevski2017}. The wireless channel is assumed to be symmetric. In networking scenarios with asymmetric wireless links, the lowest capacity among the two directions should be considered. The maximum capacity of the wireless channel is equal to $C^{MAX}$ bit/s. The wireless channel is shared and we assume that every $\text{{UAV}}_i$ can listen to any other $\text{{UAV}}_j$, including $\text{{UAV}}_0$. This is a realistic assumption, since a flying network is expected to provide wireless connectivity over a confined area of a few hundred square meters (e.g., disaster area or music festival venue) using IEEE 802.11 or 4G/5G communications technologies, which provide a high enough communications range, while benefiting from an improved Line of Sight component when compared with terrestrial networks. The Carrier Sense Multiple Access with Collision Avoidance (CSMA/CA) mechanism is employed for Medium Access Control, which is in charge of avoiding collisions of frames by enabling transmissions only when the wireless channel is sensed to be idle.

\begin{subequations}
	\begin{alignat}{10}
		& \underset{P_T, (x_0, y_0, z_0), Q_i}{\textrm{minimize}} && C(t_k)=\sum_{i=1}^{N-1}C_{0,i}(t_k)\label{eq:objective-function} \\
		& \textrm{subject to:} \notag \\
		&   && \hspace{-5em} 0 \leq P_T \leq P_T^{MAX}\label{eq:constraint1} \\
		&   && \hspace{-5em} C(t_k) \leq C^{MAX}\label{eq:constraint2} \\
		  &   &   & \hspace{-5em} 0 < T_i(t_k) \leq C_{0,i}(t_k), & \hspace{-4em} i \in \{1, ..., N-1\}\label{eq:constraint3}  \\
		  &   &   & \hspace{-5em} Q_i(t_k) \geq 0,                & \hspace{-4em} i \in \{1, ...,  N-1\}\label{eq:constraint4} \\
		  &   &   & \hspace{-5em} D_i(t_k) < H,                   & \hspace{-4em} i \in\{1, ..., N-1\}\label{eq:constraint5}   \\
		  &   &   & \hspace{-5em} 0 \leq x_i \leq x^{MAX},        & \hspace{-4em} i \in \{0, ..., N-1\}\label{eq:constraint6}  \\
		  &   &   & \hspace{-5em} 0 \leq y_i \leq y^{MAX},        & \hspace{-4em} i \in \{0, ..., N-1\}\label{eq:constraint7}  \\
		  &   &   & \hspace{-5em} 0 \leq z_i \leq z^{MAX},        & \hspace{-4em} i \in \{0, ..., N-1\}\label{eq:constraint8}  \\
		  &   &   & \hspace{-5em} (0, i), (i, 0) \in E(t_k),      & \hspace{-4em} i \in \{1, ..., N-1\} \label{eq:constraint9} \\
		&   && \hspace{-5em} (x_0 - x_i)^2 + (y_0 - y_i)^2 \notag \\ 
		  &   &   & \hspace{-5em} + (z_0 - z_i)^2 > M^2,          & \hspace{-4em} i \in\{1, ..., N-1\}\label{eq:constraint10}  
	\end{alignat}
	\label{eq:optimization-problem}
\end{subequations}

%%% PROBLEM
\textbf{Problem:} Considering $N-1$ FAPs, each generating a traffic flow $F_{0,i}$ with bitrate $T_i(t_k)$~\SI{}{bit/s} towards $\text{{UAV}}_0$, we aim at determining, at each periodic time instant $t_k$, the position of $\text{{UAV}}_0$, $P_0 = (x_0, y_0, z_0)$, the transmission power of each $\text{FAP}_i$, $P_T$, considering $P_T^{MAX}$ as the maximum transmission power allowed for the wireless technology used, and the queue size for each $\text{FAP}_i$, $Q_i$, so that: 1) $C_{0,i}(t_k)$ is high enough for accommodating $T_i(t_k)$~\SI{}{bit/s} and $C(t_k) = \sum_{i=1}^{N-1}C_{0,i}(t_k)$ is minimized; 2) the average delay $D_i(t_k)$ is below a given threshold $H$.

Our objective function, which minimizes the overall capacity required to carry the traffic offered by the FAPs, is defined in~\eqref{eq:objective-function}. Using the minimum capacity $C_{0,i}(t_k)$ aims at reducing the number of communications resources required to carry $T_i(t_k)$~\SI{}{bit/s}. In our optimization problem, we employ the transmission power $P_T$ as the fine-tuning communications resource. However, this rationale is valid for any other communications resource and corresponding optimization problem formulation, such as when considering the channel bandwidth $B$ as the fine-tuning parameter for computing the capacity $C_{0,i}(t_k)$ by means of the Shannon–Hartley theorem~\cite{Hartley1928}. A discussion on the quality of the admissible solutions obtained, considering a reference networking scenario, is presented in~\ref{appendix:a}.

In the optimization problem formulated in \cref{eq:optimization-problem}, the following constraints are considered for any time instant $t_k$:
\begin{itemize}
	\item \cref{eq:constraint1} guarantees that the computed transmission power $P_T$ is between 0 and the maximum value $P_T^{MAX}$ allowed for the wireless technology used.
	\item \cref{eq:constraint2} ensures that the aggregate capacity $C(t_k)$ of the wireless links established with $\text{{UAV}}_0$ is lower than or equal to the maximum capacity $C^{MAX}$ of the shared wireless channel used.
	\item \cref{eq:constraint3} assures that the capacity of the wireless link established between $\text{{UAV}}_i$ and $\text{{UAV}}_0$ is higher than or equal to the bitrate of the traffic flow transmitted by $\text{{UAV}}_i$.
	\item \cref{eq:constraint4} guarantees that the computed queue size for $\text{{UAV}}_i$ is not negative.
	\item \cref{eq:constraint5} ensures that the average delay of the packets generated by $\text{{UAV}}_i$ is below a given threshold $H$, which is defined according to the QoS levels to be offered by the flying network.
	\item \cref{eq:constraint6}, \cref{eq:constraint7}, \cref{eq:constraint8} ensure that the position $P_i = (x_i, y_i, z_i)$ of $\text{{UAV}}_i$ is within the dimensions $x^{MAX}$, $y^{MAX}$, and $z^{MAX}$ that characterize the 3D space where the flying network should be deployed. 
	\item \cref{eq:constraint9} assures that a wireless link capable of being established between $\text{{UAV}}_0$ and $\text{{UAV}}_i$ is always available.
	\item \cref{eq:constraint10} guarantees that the position computed for $\text{{UAV}}_0$ is $M$ meters away from the position of any $\text{{UAV}}_i$ composing the flying network, in order to avoid collision between UAVs. 
\end{itemize}

When it comes to the 3D space where the UAVs should be deployed, in the current version of our problem formulation we generically consider that the position $P_i=(x_i, y_i, z_i)$ of each $\text{{UAV}}_i$ is within zero and maximum values $x^{MAX}$, $y^{MAX}$, and $z^{MAX}$, which characterize the dimensions of the 3D space available. For safety reasons in a real-world deployment, the minimum value considered for the altitude $z_i$ of $\text{{UAV}}_i$ should assume a minimum safety value greater than \SI{0}{\meter}, so that it is positioned above the height of any ground user. This should be defined in constraint \cref{eq:constraint6}. Likewise, in order to avoid collision with any obstacles existing at a given location, the 3D space available for placing the UAVs can be defined by means of the constraints \cref{eq:constraint6}, \cref{eq:constraint7}, and \cref{eq:constraint8}. Although ensuring safety when placing UAVs is a relevant research topic, it is beyond the scope of this article. Constraints regarding the energy consumption of the UAVs are also not considered in this article. This research challenge is approached in~\cite{rodrigues2022-1, rodrigues2022-2}. 

Solving optimization problems typically requires overcoming many challenges, including non-differentiable functions and constraints, large dimensionality, and multiple local minima \cite{stanford2012}. Existing solvers can be divided into two main groups: gradient-based and gradient-free solvers. Gradient-based solvers are best-suited at finding local minima for high-dimension, non-linear constrained, and convex optimization problems. However, they are not adequate for non-convex optimization problems. In addition, their computational complexity typically grows exponentially with the problem size \cite{kulkarni2010}. Gradient-free solvers are better suited for non-convex problems. Nevertheless, contrary to gradient-based solvers, gradient-free solvers do not guarantee optimal solutions, but only acceptable solutions. Moreover, the variables that influence the values of $P_T$, $P_0 = (x_0, y_0, z_0)$, and $Q_i$ in the problem formulated in \cref{eq:optimization-problem} include: 1) the capacity of the wireless links between each $\text{FAP}_i$ and $\text{{UAV}}_0$ (FGW), at time $t_k$; 2) the number of UAVs that use the same wireless channel; 3) the behavior of the Medium Access Control protocol; and 4) the interference between UAVs. These factors are difficult to model mathematically.

The objective function in our problem formulation aims at minimizing the overall capacity $C(t_k)$ of the wireless links established between all the FAPs and $\text{{UAV}}_0$ (FGW) at each instant $t_k$. The wireless links' capacity is expressed by means of the Shannon-Hartley theorem. The Shannon-Hartley theorem allows to compute the maximum amount of information that can be carried per unit of time by a communications channel subject to a certain $\text{{SNR}}_i$, using a given bandwidth $B$, as follows: $C = B \times log_2(1 + \text{{SNR}}_i)$. This equation is a non-convex function~\cite{Tychogiorgos2011, Tychogiorgos2013}. Since the objective function consists of the sum of multiple non-convex functions, each representing the capacity for the wireless link established between a FAP and the FGW computed using the Shannon-Hartley theorem, then it is a non-convex function too. The proof is presented in~\cref{eq:non_convex_proof}.

\begin{multline}
	\label{eq:non_convex_proof}
	C=B\times\log_2\left(1+SNR_1\right)+B\times\log_2\left(1+SNR_2\right)+\\\ldots+B\times\log_2\left(1+SNR_{N-1}\right)=B\times\log_2\big[\left(1+SNR_1\right)\\\times\left(1+SNR_2\right)\times\ldots\times\left(1+SNR_{N-1}\right)\big]=B\times\log_2\left(u\right)
\end{multline}

where:
\begin{description}
	\item[u=\prod_{i=1}^{N-1}\left(1+SNR_i\right)]
\end{description}

The non-convex nature of the objective function prevents any gradient-based algorithm from converging to a global optimal solution.

\begin{figure}[!t]
	\centering
	\includegraphics[width=1\linewidth]{"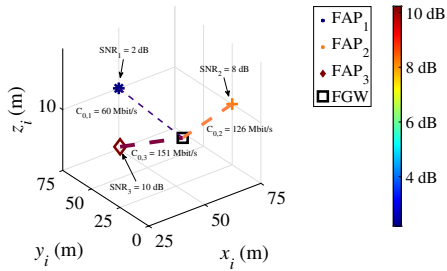"}
	\caption{Theoretical SNR and capacity values of the wireless links established between each FAP and the FGW, obtained by solving the optimization problem defined in~\cref{eq:optimization-problem}. The colors of the markers that represent the FAPs are associated with the order of magnitude of the SNR values of the wireless links established with the FGW, considering the depicted color bar.}
	\label{fig:theoretical_results}
\end{figure}

For the sake of theoretical analysis, we solved the optimization problem using the GEKKO optimization suite \cite{beal2018gekko}. For that purpose, we took into account the Shannon-Hartley theorem~\cite{Hartley1928} for determining the capacity $C_{0,i}(t_k)$ of the wireless link established between each $\text{{FAP}}_i$ and $\text{{UAV}}_0$, while the queue of each FAP was modeled using the M/D/1 queuing model~\cite{bershkas}, considering a constant packet size $S$ equal to \SI{1400}{bytes}. $\text{{SNR}}_i$ was computed by means of the Friis propagation loss model, for channel bandwidth $B$ equal to \SI{160}{\mega\hertz}, \SI{5250}{\mega\hertz} as the carrier frequency $f$, and constant noise power $P_N$ equal to \SI{-85}{dBm}. This allowed us to obtain the point $(x_0,y_0,z_0)~=~(46.4, 12.3, 10)$~\SI{}{m} for the FGW placement, UAVs' transmission power $P_{T} = $  \SI{0}{dBm}, and queue size $Q_1$, $Q_2$, and $Q_3$ equal to respectively 1, 111, and 133 packets. The SNR and capacity of the wireless links obtained are presented in~\cref{fig:theoretical_results}.

Due to the non-convex nature of the objective function, it is known that the solution obtained can be a local optimum only. Moreover, since the Shannon-Hartley theorem only provides a theoretical maximum value for the capacity of the wireless links, the capacity obtained in practice in IEEE 802.11 links is substantially lower, especially due to the influence of neighboring nodes competing to access the medium and the overhead associated with the Medium Access Control protocol. For these reasons, the obtained theoretical results lead to target SNR values lower than the ones required in practice to achieve the same channel capacity, when considering equal channel bandwidth and noise power. This can be concluded by comparing them with the minimum SNR values required for using the IEEE 802.11ac MCS indexes with a data rate close to the FAPs' traffic demand, which are presented in~\cref{tab:mapping-snr-channel-capacity}. Furthermore, the time required to obtain a local optimal or quasi-optimal solution using a state of the art solver raises exponentially as both the number of UAVs and dimensions of the 3D space where the UAVs can be deployed increase. This may avoid obtaining timely solutions in dynamic flying networks that need to be reconfigured in short periods of time. For these reasons, we propose a heuristic algorithm that takes into account domain knowledge to find an adequate traffic-aware solution for the problem formulated in \cref{eq:optimization-problem}. 

\begin{table}[!t]
	\caption{Minimum SNR required to select the IEEE 802.11ac MCS indexes 2, 5, and 7, and corresponding fair share for 3 FAPs.}
	\label{tab:mapping-snr-channel-capacity}
	\centering
	\begin{tabular}{l l l l}
		\hline
		SNR  &
		MCS &
		Data Rate &
		Fair share\\
		(dB) &
		index &
		(Mbit/s) &
		(Mbit/s)\\
		\hline
		\\
		\mymk{15}      & 2 & 175.5 & $0.80 \times \frac{175.5}{3 \text{ FAPs}} = 50$ \\
		\\
		\rectangle{27} & 5 & 468   & $0.80 \times \frac{468}{3 \text{ FAPs}} = 133$  \\
		\\
		\circled{35}   & 7 & 585   & $0.80 \times \frac{585}{3 \text{ FAPs}} = 166$  \\
		\\
		\hline
	\end{tabular}
\end{table}
%%%%%%%%%%%%%%%%%%%%%%%%%%%%%%%%%%%%%%%%%%%%%
% GPQM ALGORITHM
%%%%%%%%%%%%%%%%%%%%%%%%%%%%%%%%%%%%%%%%%%%%%
\section{Traffic-aware Gateway Placement and Queue Management (GPQM) Algorithm \label{sec:GPQM}}

The GPQM algorithm results from the combination of the GWP and PQM algorithms that we have originally proposed in respectively \cite{Coelho2019GWP} and \cite{Coelho2020}. 
On the one hand, the PQM algorithm aims at maximizing the throughput of the traffic carried between the FAPs and the FGW, while meeting an average delay constraint. Since PQM configures only the queue size of the communications nodes, it is unable to find a suitable solution that meets the average delay constraint for all flows offered by the FAPs, if the capacity of the wireless links is not high enough to achieve that. On the other hand, the GWP algorithm aims at ensuring that the wireless link established between each FAP and the FGW has high enough capacity for carrying the offered traffic. For that purpose, the GWP algorithm defines the FGW placement and the transmission power of the UAVs. However, it does not address the queue size management of the communications nodes, which may imply long queueing delays when the queue size is not adequate for the capacity of the wireless links available. As such, GWP is not able to provide average delay guarantees. Building upon the conclusion that the usage of static, oversized queues in IEEE 802.11 networks leads to underutilized communications nodes and unnecessary high delays without improving the aggregate throughput~\cite{Li2010buffer}, GPQM defines: 1) the FGW position, in order to enable a wireless link between each $\text{FAP}_i$ and the FGW that meets the traffic demand of $\text{FAP}_i$; and 2) calculating the queue size for each $\text{FAP}_i$, in order to provide stochastic delay guarantees. 
The GPQM algorithm allows packets to be held in the queue of each $\text{FAP}_i$ for a short time, while maximizing the aggregate throughput measured in the FGW and improving the shared channel usage. Overall, this allows to improve the performance of the flying network. In what follows, we explain the GPQM algorithm, which is presented in~\cref{alg:gpqm-algorithm}.

\begin{algorithm}[!t]
	\caption{GPQM algorithm}\label{alg:gpqm-algorithm}
	\DontPrintSemicolon
	\KwInput{\scriptsize \textbf{Traffic demand} \bm{$T_i$}, in bit/s, of each $\text{FAP}_i$, \textbf{packet size} \bm{$S$} in bytes, and \textbf{position} \bm{$P_i$}, in 3D Cartesian coordinates, of each $\text{FAP}_i$}
	\KwOutput{\scriptsize \textbf{Transmission (TX) power} \bm{$P_T$}, in dBm, for all UAVs, \textbf{FGW position} \bm{$P_0$} in 3D Cartesian coordinates, and \textbf{queue size} \bm{$Q_i$}, in number of packets, for each $\text{FAP}_i$}
	\tcc{\scriptsize \SI{0}{dBm} TX power}
	$P_T \gets 0$\\
	\tcc{\scriptsize $\text{FAP}_i$ packet inter-arrival time}
	$\lambda_i \gets \frac{T_i}{S \times \SI{8}{bits/packet}}$\\
	\tcc{\scriptsize Iterate up to maximum allowed TX power -> O(n)}
	\While{$P_T \leq P_T^{MAX}$}
	{
		\tcc{\scriptsize Compute the FGW position, considering $i\in\{1, ..., N-1\}$. $\text{SNR}_i$ is determined taking into account $T_i$ by means of a lookup table similar to~\cref{tab:mapping-snr-channel-capacity}}
		computeFgwPos($[x_i]$, $[y_i]$, $[z_i]$, $[\text{SNR}_i]$, $P_T$)\\
		\tcc{\scriptsize i.e., $(x_0, y_0, z_0) \in S_G$}
		\If{$(x_0, y_0, z_0) \neq \oslash$}
		{
			\tcc{\scriptsize $\text{FAP}_i$ packet service time}
			$\mu_i \gets \frac{C_{0,i}}{S \times \SI{8}{bits/packet}}$\\
			\tcc{\scriptsize $\text{FAP}_i$ traffic load}
			$\rho_i = \frac{\lambda_i}{\mu_i}$\\
			\tcc{\scriptsize $\text{FAP}_i$ queue size}
			$Q_i = \frac{1}{2} \times \frac{\rho_i^2}{1-\rho_i}$\\
			\tcc{\scriptsize $\text{FAP}_i$ delay}
			$D_i = \frac{2-\rho_i}{2 \times \mu_i \times (1-\rho_i)}$\\
			\tcc{\scriptsize $\text{FAP}_i$ delay lower than max. delay $H$}
			\If{$D_i < H$}
			{
				\tcc{\scriptsize TX power for all UAVs, FGW position, and queue size for each $\text{FAP}_i$}
				\Return $P_T, (x_0, y_0, z_0), Q_i$
			}
		}
		\Else
		{
			\tcc{\scriptsize Increase TX power by \SI{1}{dBm}}
			$P_T \gets P_T + 1$
		}
	}
	\tcc{\scriptsize Compute and return $x_0$, $y_0$, and $z_0$ -> O(1)}
	\SetKwFunction{FMain}{computeFgwPos}
	\SetKwProg{Fn}{Function}{:}{}
	\Fn{\FMain{$[x_i]$, $[y_i]$, $[z_i]$, $[\text{SNR}_i]$, $P_T$}}{
		\tcc{\scriptsize System of equations in the form of~\cref{eq:gw-placement-equations-system}}
		\scriptsize$(x_0-x_1)^2 + (y_0-y_1)^2 + (z_0-z_1)^2 \leqslant \qquad \left(10^{\frac{K + P_T-\text{SNR}_1}{20}}\right)^2$ \;
		...\;
		$(x_0-x_{N-1})^2 + (y_0-y_{N-1})^2 + (z_0-z_{N-1})^2 \leqslant
		\left(10^{\frac{K + \scriptsize P_T-\text{SNR}_{N-1}}{20}}\right)^2$ \;
		$K = -20\log_{10}(f_i) - 20\log_{10}\left(\frac{4\times\pi}{c}\right) - (P_N)$\;
		\KwRet {($x_0$, $y_0$, $z_0$)}\;
	}
\end{algorithm}

GPQM takes advantage of the centralized view of the flying network available at the Central Station, where the flying network is controlled. For that purpose, the Central Station takes into account the information collected and transmitted periodically by the FAPs themselves, including the traffic demand and the positions of the ground users. Based on this information, the FAP placement algorithm, which is beyond the scope of this article, determines the FAPs' positions and the cell ranges of the access network, in order to meet the traffic demand of the ground users. Before transmitting the updated information to the FAPs, the GPQM algorithm calculates the FGW position and the queue size for each FAP. 
For the sake of simplicity, we omit $t_k$ in what follows. Considering the future positions of $\text{FAP}_i$ and the bitrate of the traffic flow $F_{0,i}$, $T_i$~\SI{}{bit/s}, GPQM aims at guaranteeing that a wireless link available between $\text{FAP}_i$ and $\text{{UAV}}_0$ (FGW) has a minimum $\text{SNR}_i$ that enables the usage of a $\text{{MCS}}_i$ index, characterized by a physical data rate higher than or equal to $T_i$~\SI{}{bit/s}, and allows an average delay $D_i$ lower than $H$ seconds. Conceptually, if $\text{{MCS}}_i$ is ensured, then the network will be able to accommodate $T_i$~\SI{}{bit/s}, allowing to maximize the amount of bits received in the FGW and to minimize the average delay $D_i$. 

The selection of $\text{{MCS}}_i$ imposes a minimum $\text{SNR}_i$, considering a constant noise power $P_N$. If the transmission power $P_{T}$ is known, we can calculate the maximum admissible distance $d_{max_i}$ between $\text{FAP}_i$ and $\text{{UAV}}_0$, using the Friis propagation loss model defined in~\eqref{eq:friis-propagation-model} in \SI{}{\decibel}, where $f$ is the carrier frequency, $\pi$ is the mathematical constant defined in Euclidean geometry as the ratio of a circle's circumference to its diameter, and $c$ represents the speed of light in vacuum. 

\begin{equation}
	\begin{aligned}
		\text{SNR}_i = P_{T} - 20\log_{10}(d_{max_i}) - 20\log_{10}(f) - \\ 20\log_{10} \bigl(\begin{smallmatrix}
		\frac{4 \times \pi}{c}                                           
		\end{smallmatrix}\bigr)                                          
		-P_N                                                             
	\end{aligned}
	\label{eq:friis-propagation-model}
\end{equation}

\begin{figure}[!t]
	\centering
	\includegraphics[width=1\linewidth]{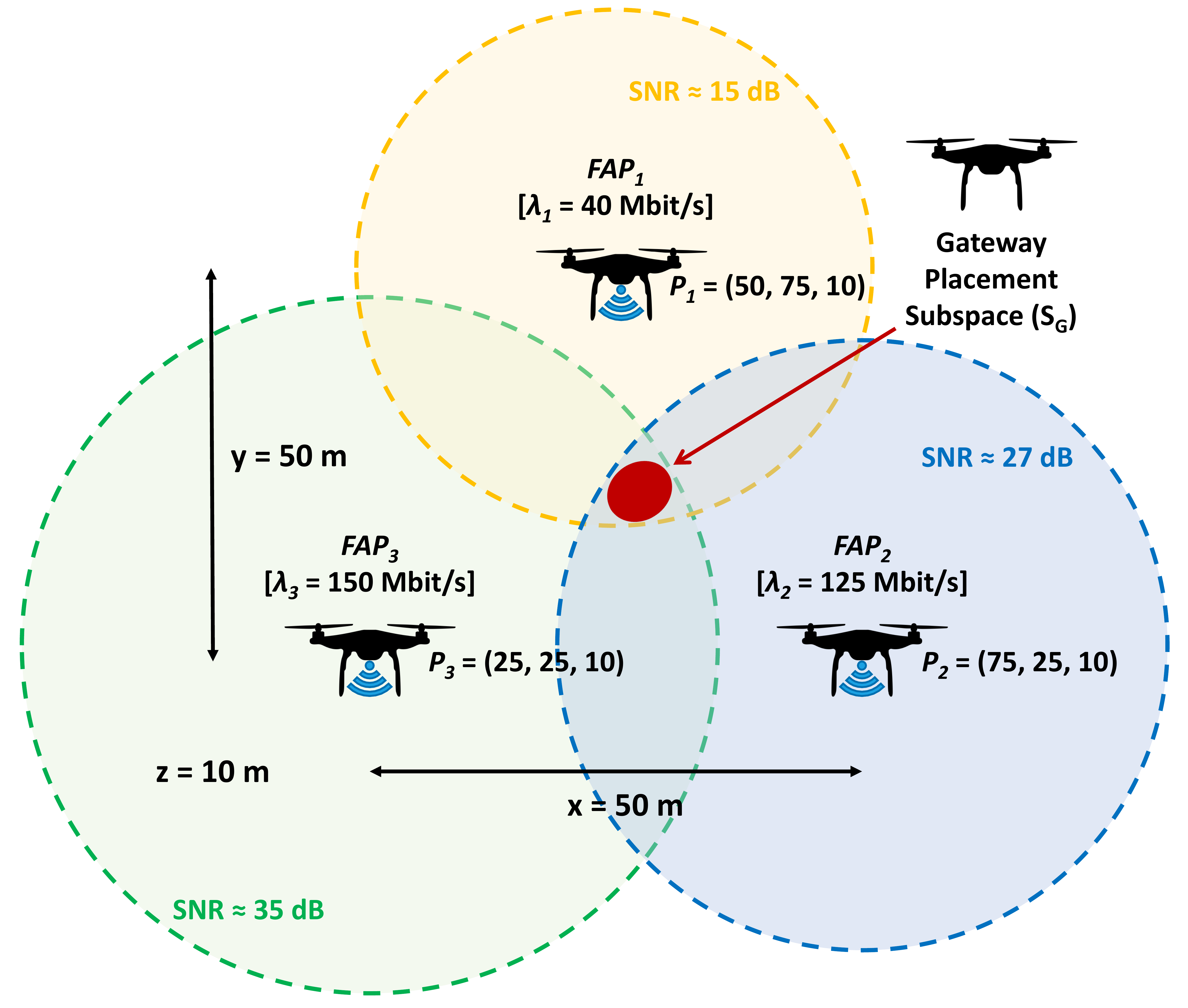}
	\caption{Reference scenario depicting the Gateway Placement Subspace ($S_G$) in a two-Dimensional (2D) space, which results from the intersection of the circumferences, centered at each FAP, with radius equal to the maximum distance compliant with the minimum SNR required.}
	\label{fig:reference-scenario}
\end{figure}

In 3D space, $d_{max_i}$ corresponds to the radius of a sphere, centered at $\text{FAP}_i$, inside which $\text{{UAV}}_0$ should be placed. Considering $N-1$ UAVs, the placement subspace for positioning $\text{{UAV}}_0$ is defined by the intersection of the corresponding spheres $i \in \{1, ..., N-1\}$; we refer to this subspace as the Gateway Placement Subspace, $S_G$, as depicted in \cref{fig:reference-scenario}. $S_G$ is calculated using GPQM, which iteratively allows obtaining the point $P_0 = (x_0, y_0, z_0)$ for positioning $\text{{UAV}}_0$ and the minimum transmission power $P_T$ that is assumed to be the same for all UAVs. If no solution is found, \cref{alg:gpqm-algorithm} is terminated, since it constitutes an access network planning problem; in order to achieve a solution for the FGW placement, additional FAPs or network relays should be included in the flying network, aiming at enabling shorter wireless links established between UAVs configured with transmission power $P_T$ lower than or equal to $P_T^{MAX}$.

%\subsection{Proactive Queue Management \label{sec:PQM}}
In addition, the GPQM algorithm ensures that $\text{FAP}_i$ is configured, at each periodic time instant $t_k$, with a queue size $Q_i$ able to accommodate the traffic sent towards $\text{{UAV}}_0$. We assume the M/D/1 queuing model, according to the Kendall's notation \cite{bershkas}, considering the data packets generated by the users being served by the FAPs have constant size $S$ and mutually independent inter-arrival times. The packet arrival at each $\text{FAP}_i$ is assumed to follow a Poisson process. The inter-arrival times follow an exponential distribution whose average is $\lambda_i$ \SI{}{packet/s}. The packet service time is defined by $\mu_i = C_{0,i}/(S \times 8)$ \SI{}{packet/s}, while the traffic load of each FAP is expressed by $\rho_i = \lambda_i/\mu_i$. The average number of packets $Q_i$ in each FAP's queue is calculated using \cref{eq:queue-size} and the average packet delay is calculated using \cref{eq:waiting-time} \cite{bershkas}, which includes the waiting time in the queue plus the packet service time. By enabling $\text{FAP}_i$ to have a network queue of size $Q_i$, we aim at providing an average delay $D_i$ below a given threshold $H$ during $t_k$. 

\begin{equation}
	Q_i = \frac{1}{2} \times \frac{\rho_i^2}{1-\rho_i}
	\label{eq:queue-size}
\end{equation}

\begin{equation}
	D_i = \frac{2-\rho_i}{2 \times \mu_i \times (1-\rho_i)}
	\label{eq:waiting-time}
\end{equation}

The pseudocode presented in~\cref{alg:gpqm-algorithm} is considered as reference for analyzing the time complexity of the GPQM algorithm. Considering the partial time complexities presented above line 3 and line 14, namely $O(n)$ and $O(1)$, we conclude the GPQM algorithm has time complexity $O(n)$. $n$ represents the transmission power levels required for achieving a valid solution for the problem. $P_T$, initially set to \SI{0}{dBm}, is successively increased by \SI{1}{dBm} up to $n$ times, until a valid solution is found. 
	
The GPQM algorithm always converges to a valid solution, since it is a heuristic algorithm leveraged on the theoretical domain knowledge that improving the transmission power $P_T$ of the FAPs leads to higher SNR and improved network capacity for the wireless links. As such, the spheres that represent the transmission ranges of the FAPs always intersect between themselves at a certain transmission power level $P_T$, unless the maximum transmission power level $P_T^{MAX}$ allowed for the wireless technology used is reached, which is not a limitation of the algorithm itself.

\begin{figure}[!ht]
	\centering
	\includegraphics[width=1\linewidth]{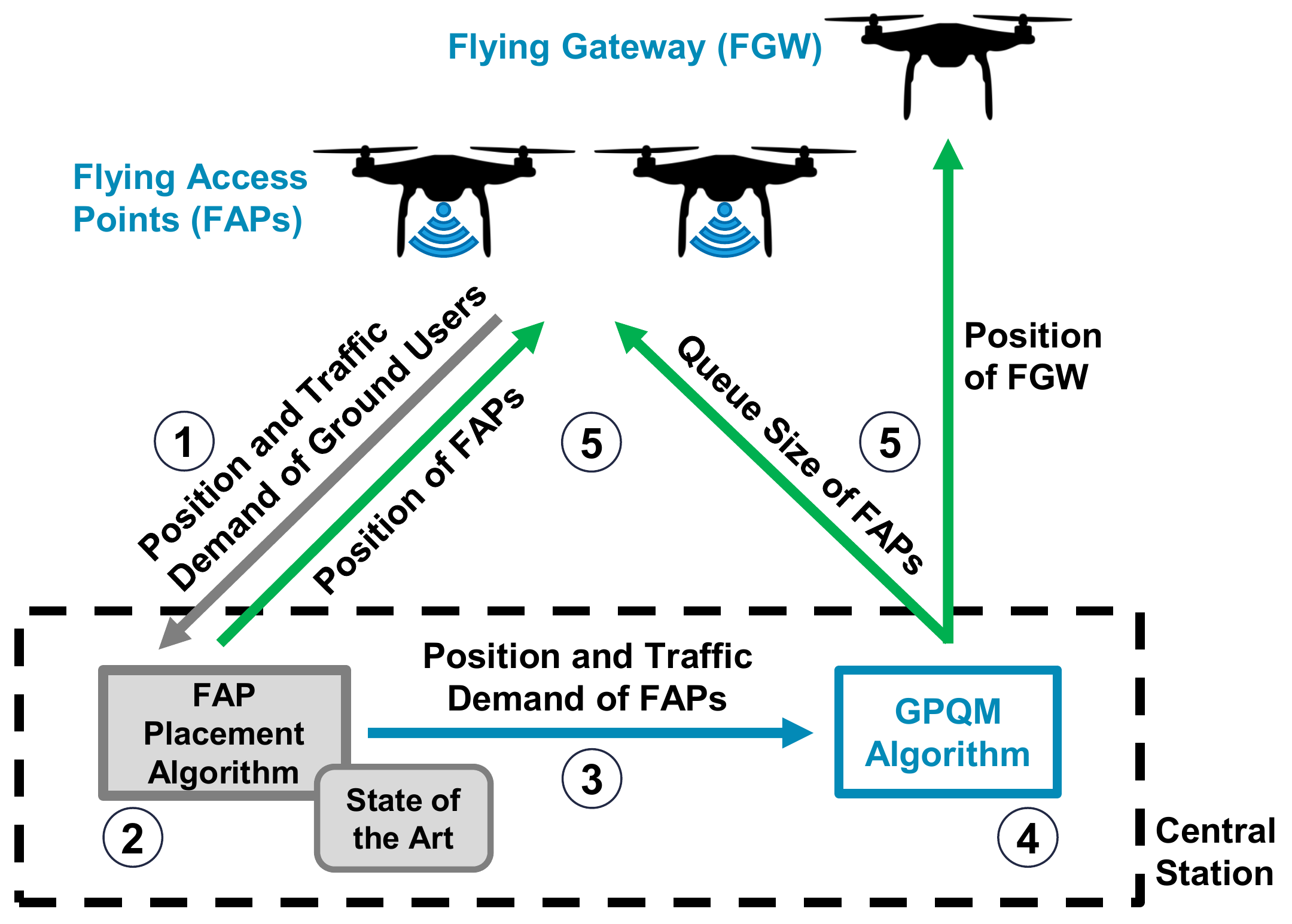}
	\caption{Information exchanged between the flying network, the FAP placement algorithm, and the GPQM algorithm. The circled numbers from 1 to 5 define the sequential order considered. The flying network, including the UAVs positions and queue size, is reconfigured with period $\Delta t \gg \SI{1}{\second}$.}
	\label{fig:system-elements}
\end{figure}

The information exchanged between the flying network, the FAP placement algorithm, and the GPQM algorithm is depicted in~\cref{fig:system-elements}, including the sequential order considered in each flying network reconfiguration.

\subsection{Illustrative Example}\label{sec:IllustrativeExample}
Herein, we exemplify the execution of the GPQM algorithm for the simple scenario shown in \cref{fig:reference-scenario}. It is composed of three FAPs that are placed in a 50 by 50 meters square, hovering at 10 meters altitude. In this illustrative example, we consider the use of the IEEE 802.11ac technology with one spatial stream, \SI{160}{\mega\hertz} channel bandwidth, and \SI{800}{\nano\second} Guard Interval. Let us assume that the demanded capacity values for $\text{FAP}_1$, $\text{FAP}_2$, and $\text{FAP}_3$ are, respectively, \SI{175.5}{Mbit/s}, \SI{468}{Mbit/s}, and \SI{585}{Mbit/s}, which correspond respectively to the MCS indexes 2, 5, and 7. These demanded capacities were selected in order to exemplify the traffic-awareness of the GPQM algorithm when the FAPs have different traffic demand values, which may occur in the real-world due to different number of ground users served. Taking into account that a minimum SNR value is required to use a target MCS index, which is characterized by a theoretical data rate, in \cref{tab:mapping-snr-channel-capacity} we conclude that the target SNR values are \SI{15}{\deci\bel} for $\text{FAP}_1$, \SI{27}{\deci\bel} for $\text{FAP}_2$, and \SI{35}{\deci\bel} for $\text{FAP}_3$.

Solving the system of equations~\eqref{eq:gw-placement-equations-system}, which is derived from~\eqref{eq:friis-propagation-model}, and considering $d_{max_i}$ as the Euclidean distance between $\text{{UAV}}_0$ (FGW) and each $\text{FAP}_i$, we conclude that a suitable position for the FGW is $(x_0,y_0,z_0)~=~(39.0, 21.9, 9.2)$~\SI{}{m}, for a transmission power $P_{T} = $  \SI{20}{dBm}. In the system of equations~\eqref{eq:gw-placement-equations-system}, $P_T$ is the fine-tuning parameter, so that we can find at least a point $(x_0, y_0, z_0) \in S_G$. $P_T$ is initially set to \SI{0}{dBm}. Then, it is iteratively increased by \SI{1}{dBm} until a valid solution is found or the maximum transmission power allowed $P_T^{MAX}$ is reached. The factor $K$ is calculated considering the carrier frequency $f$ used and the constant noise power $P_N$: \SI{5250}{\mega\hertz} and \SI{-85}{dBm}, respectively. 

\begin{equation} \label{eq:gw-placement-equations-system}
	\left\{
	\begin{aligned}
		  & (x_0-50)^2 + (y_0-75)^2 + (z_0-10)^2 \leqslant                        \\
		  & \qquad \left(10^{\frac{K + P_T-\mymk{15}}{20}}\right)^2               \\
		  & (x_0-75)^2 + (y_0-25)^2 + (z_0-10)^2 \leqslant                        \\ 
		  & \qquad \left(10^{\frac{K+P_T-\rectangle{27}}{20}} \right)^2           \\
		  & (x_0-25)^2 + (y_0-25)^2 + (z_0-10)^2 \leqslant                        \\
		  & \qquad \left(10^{\frac{K+P_T-\circled{35}}{20}} \right)^2             \\
		  & K = -20\log_{10}(5250\times10^6) -                                    \\
		  & \qquad 20\log_{10}\left(\frac{4\times\pi}{3\times10^8}\right) - (-85) \\
	\end{aligned}
	\right.
\end{equation}

Since the wireless channel is shared by $N-1$ FAPs generating traffic towards $\text{{UAV}}_0$ (FGW), in order to define the bitrate of the traffic flow transmitted by the FAPs, we assumed a reference fair share $L = 0.80 \times (\SI{585}{Mbit/s}) / (N-1)$ of the maximum channel capacity allowed, where $\eta = 0.80$ is the efficiency of the Medium Access Control protocol, including the overhead of the transport protocol, Internet Protocol (IP), and Medium Access Control packet headers. \SI{585}{Mbit/s} is the data rate associated to the MCS index 7, which is the highest MCS index to be ensured by the flying network so that the traffic demand of $\text{FAP}_3$ is met. The minimum SNR required to use each IEEE 802.11ac MCS index (cf.~\cref{tab:mapping-snr-channel-capacity}) and the factor $\eta$ were obtained by means of ns-3 simulations, considering a FAP generating UDP Poisson traffic towards the FGW, for a constant transmission power equal to \SI{20}{dBm}. For that purpose, the FAP was moving away from the FGW, in order to increase the Euclidean distance, leading to the degradation of the wireless link's SNR, and enabling the selection of multiple MCS indexes. This allowed establishing the relation between the wireless link's SNR values and the selected MCS indexes.

For the networking scenario depicted in \cref{fig:reference-scenario}, let us assume that the bitrate of the traffic offered by the FAPs is equal to \SI{40}{Mbit/s}, \SI{125}{Mbit/s}, and \SI{150}{Mbit/s}, which correspond, respectively, to approximately 25\%, 75\%, and 90\% of the reference fair share $L$ (\SI{166}{Mbit/s}) of the maximum channel capacity allowed (cf.~\cref{tab:mapping-snr-channel-capacity}). In order to calculate the queue size, a constant packet size $S$ equal to \SI{1400}{bytes} is considered in this example, which is close to the Maximum Transmission Unit (MTU) of \SI{1500}{bytes} typical in data packets composing the Internet traffic \cite{8845315}. Considering the specific case of $\text{FAP}_3$ for illustrative purposes, this results in $\lambda_3 = (\SI{150}{Mbit/s})/ (1400 \times \SI{8}{bits/packet}) = \SI{13393}{packet/s}$ and $\mu_3 = (\SI{166}{Mbit/s}) / (1400 \times \SI{8}{bits/packet}) = \SI{14821}{packet/s}$, leading to $\rho_3 \approx 0.9$ and $Q_3 = \SI{5}{packets}$, according to \cref{eq:queue-size}. We argue that this queue size allows that packets are held in the transmission queue for shorter time, thus decreasing the queuing delay and maximizing the throughput, when compared to the baseline that does not employ any queue management mechanism. The same rationale is employed for the remaining FAPs that compose the flying network. 

\begin{figure}[!t]
	\centering
	\includegraphics[width=0.80\linewidth]{"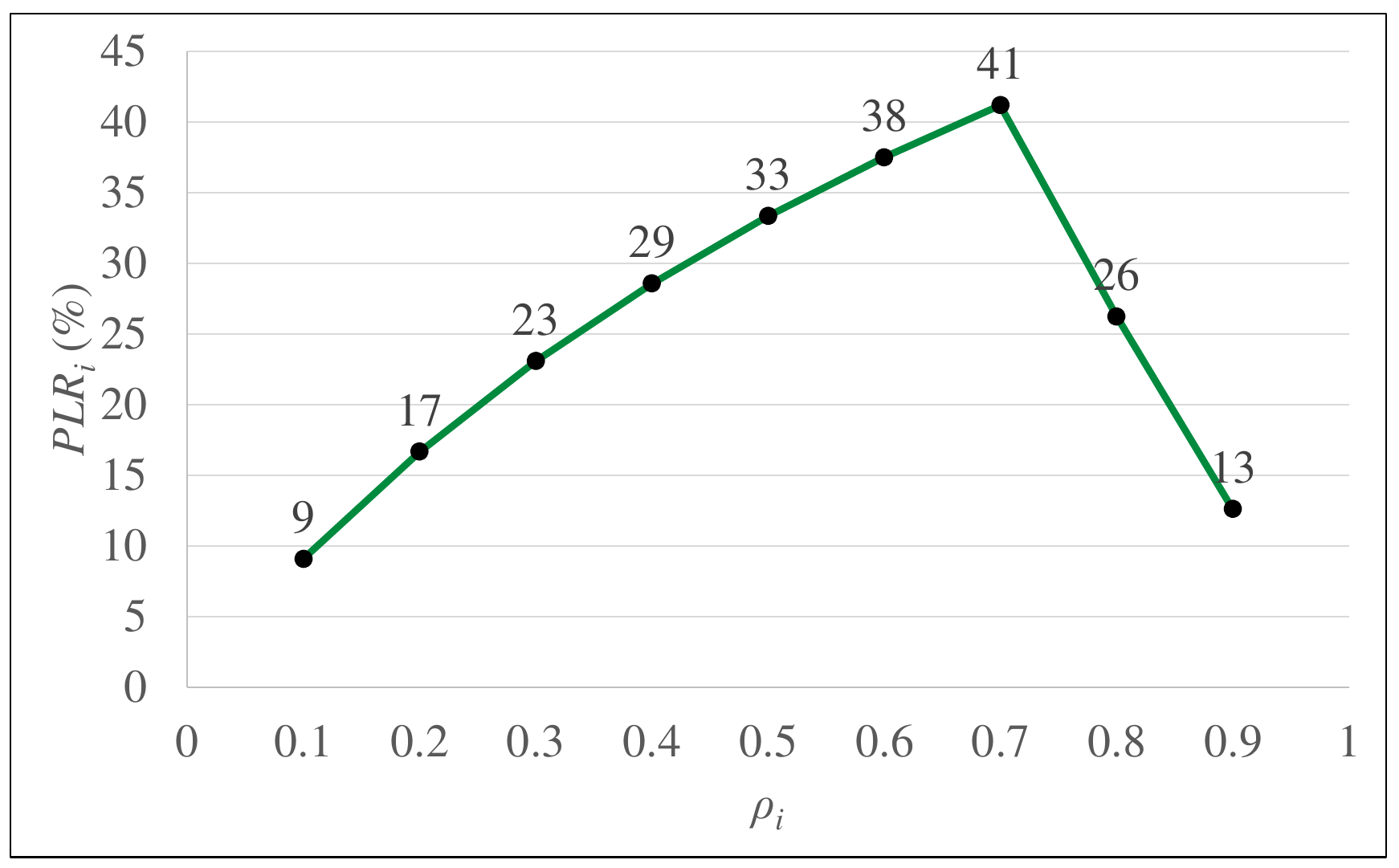"}
	\caption{Packet Loss Ratio ($PLR_i$) versus traffic load $\rho_i$.}
	\label{fig:plr-vs-rho}
\end{figure}

In \cite{Coelho2020}, a sensitivity analysis proved that, in fact, the queue size $Q_i$, obtained by means of \cref{eq:queue-size}, enables maximum aggregate throughput with minimum average delay, when compared with other possible queue size values. $Q_i$, which represents the average number of packets in the queue of $\text{FAP}_i$, considers the M/D/1 queuing model and depends directly on $\rho_i$. The Packet Loss Ratio ($PLR_i$) of $\text{FAP}_i$ for different values of $\rho_i$, when employing $Q_i$ as the queue size, can be calculated using \cref{eq:blocking-probability}. Although the queue configured in $\text{FAP}_i$ is represented by the M/D/1/$Q_i$ queuing model, for calculating $PLR_i$ we take into account the M/M/1/$Q_i$ queuing model, since the blocking probability $p_{b_i}$ does not depend on the packet service time distribution \cite{bershkas}. We conclude that $PLR_i$ is up to 41\% for $\rho_i=0.7$, as depicted in \cref{fig:plr-vs-rho}. As the traffic load $\rho_i$ associated with $\text{{FAP}}_i$ increases, the average number of packets in its queue increases exponentially. For $\rho_i$ between 0.1 and 0.7 the average number of packets $Q_i$ is 1, whereas for $\rho_i$ equal to 0.8 and 0.9 $Q_i$ is respectively equal to 2 and 4. This increase in the number of packets that the queue of $\text{{FAP}}_i$ can accommodate justifies the decrease of the $PLR_i$ depicted in \cref{fig:plr-vs-rho}.

\begin{equation}
	PLR_i = p_{b_i} = P[Q_i] = \frac{1-\rho_i}{1-\rho_i^{Q_i+1}}\times\rho_i^{Q_i}
	\label{eq:blocking-probability}
\end{equation}
  
Despite $Q_i$ enables maximum aggregate throughput with minimum average delay, the queue size of each $\text{FAP}_i$ should be fine-tuned when low $PLR_i$ is considered for service level agreements, such as to enable ultra-reliable communications. In order to reduce $PLR_i$, $Q_i$ should be increased, at the expense of a longer average delay $D_i$. On the other hand, since $\rho_i = \lambda_i / \mu_i = \lambda_i / (C_{0,i}/(S\times8))$, if the capacity $C_{0,i}$ is increased, then $\rho_i$ will be decreased, thus contributing to reduce $PLR_i$, as depicted in \cref{fig:plr-vs-rho}. This can be achieved by adjusting the FGW placement and the transmission power $P_T$, according to the rationale of GPQM, in order to ensure low delay and low PLR values simultaneously. When multiple traffic flows have different QoS requirements, multiple queues with different sizes can be used on $\text{FAP}_i$, one for each flow, following the concept proposed in \cite{taht2018}. In this sense, queues with larger size than $Q_i$ can be configured for flows that require low $PLR_i$, while queues with smaller size than $Q_i$ can be configured to achieve maximum aggregate throughput with minimum average delay for other flows. Therefore, in GPQM, the FGW placement, $P_T$, and $Q_i$ are the main fine-tuning variables to optimize the overall flying network performance. In this article, we aim at maximizing the aggregate throughput, while providing stochastic delay guarantees.

%%%%%%%%%%%%%%%%%%%%%%%%%%%%%%%%%%%%%%%%%%%%%
% PERFORMANCE EVALUATION
%%%%%%%%%%%%%%%%%%%%%%%%%%%%%%%%%%%%%%%%%%%%%
\section{Performance Evaluation \label{sec:PerformanceEvaluation}}
The performance evaluation when the GPQM algorithm is used is presented in this section. Firstly, we describe the evaluation methodology. Secondly, we characterize the simulation setup. Thirdly, we present and discuss the network performance results obtained in simulation. Finally, we compare the GPQM algorithm with state of the art solvers.

\subsection{Evaluation Methodology} \label{sec:EvaluationMethodology}
Since 1) a purely theoretical performance evaluation would require many simplifications, due to the high complexity of the problem addressed by this article, leading to unrealistic results, and 2) an experimental evaluation would be limited, with reduced scalability, and expensive, due to the inherent complexity of setting up a testbed composed of multiple UAVs, we carried out a performance evaluation based on ns-3 simulations. The ns-3 simulator \cite{ns-3} was chosen due to its accurate models for wireless networks and its wide acceptance within the scientific community.

\begin{figure}[!t]
	\centering
	\includegraphics[width=0.90\linewidth]{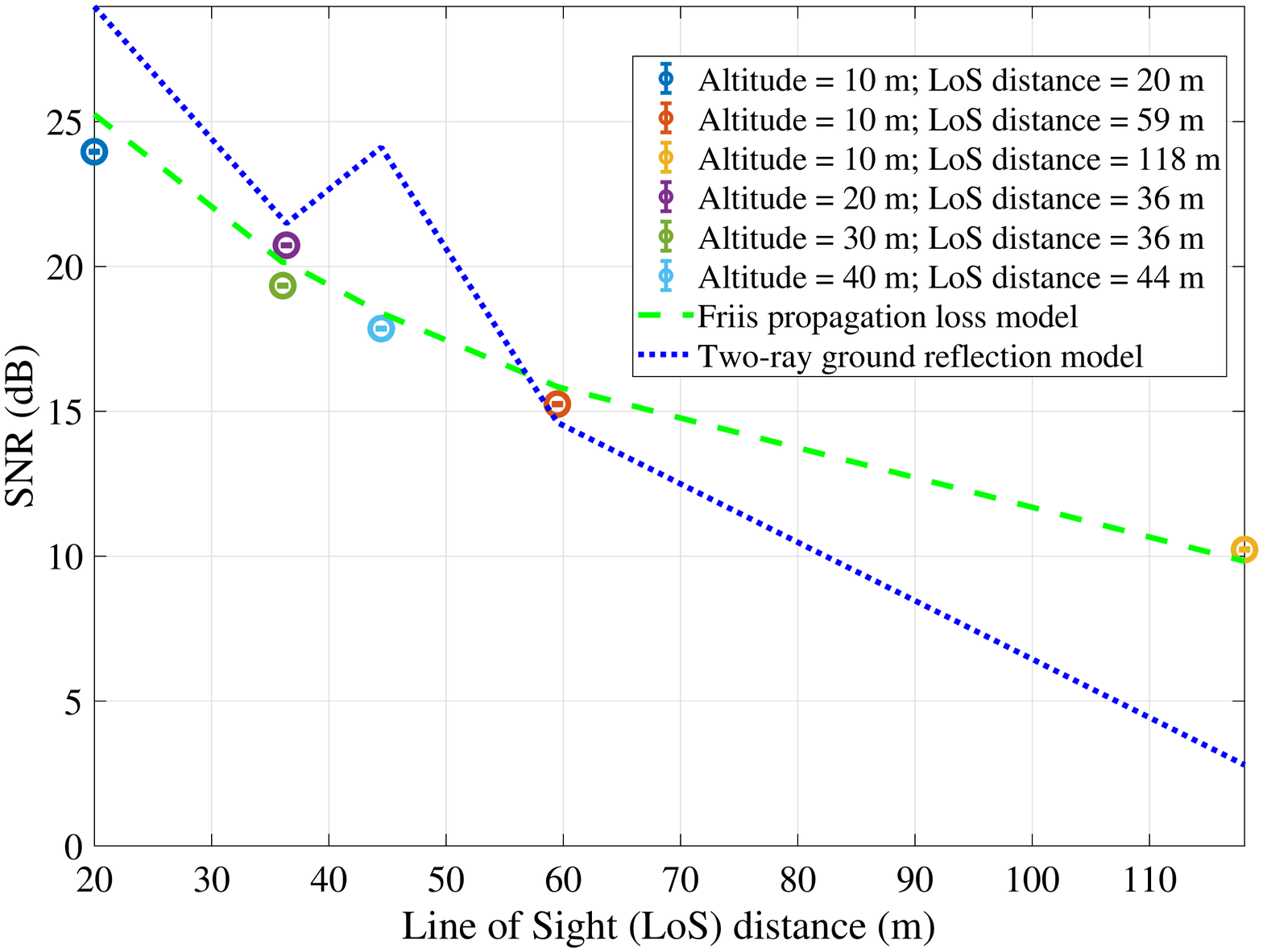}
	\caption{Experimental SNR versus the Line of Sight (LoS) distance of the wireless link, considering different UAV altitudes. The Friis propagation loss and two-ray ground reflection models are represented by the dashed and dotted lines, respectively \cite{Almeida2020}.}
	\label{fig:experimental-snr-gwp}
\end{figure}

Although the ns-3 simulator includes a set of theoretical propagation models tailored for wireless networks, in order to improve the accuracy and realism of the performance evaluation presented in this article, the wireless channel between each $\text{FAP}_i$ and the FGW was characterized based on the experimental model proposed in \cite{Almeida2020}, where it was analyzed in terms of the path-loss and fast-fading components. Regarding the path loss component, it was concluded that the Friis propagation loss model is the best suited to characterize the wireless links in flying networks. The Friis propagation loss model (cf. dashed green line in~\cref{fig:experimental-snr-gwp}) expresses that the power loss is proportional to the square of the distance between the transmitter and the receiver communications nodes, while proportional to the square of the carrier frequency used. This model fits the experimental SNR values measured in a flying network, which is justified by the strong Line of Sight component induced by the UAV altitude, as suggested in the literature~\cite{khuwaja2018}. For comparison purposes, the two-ray ground reflection model (cf. dotted blue line in~\cref{fig:experimental-snr-gwp}) is also depicted. It takes into account a component reflected on the ground and the influence of the antennas’ height, in addition to the power loss induced by the Line of Sight component. For distances between the transmitter and the receiver less than the height of the transmitter’s antenna, two waves are added in a constructive way, increasing the power of the received signal. As the distance increases, these waves add up constructively and destructively, giving rise to up-fade and down-fade lobes in the power of the received signal, as depicted in~\cref{fig:experimental-snr-gwp}.

On the other hand, the Rician fast-fading component, commonly called K-factor, which was characterized in~\cite{Almeida2020}, is also considered in our article. The K-factor constitutes the ratio of the received power in the dominant component over the scattered power. We take into consideration worst-case scenarios where the wireless links established between the UAVs are modeled by the Friis propagation loss model and the Rician fast-fading with a low K-factor value. This aims at considering the wireless conditions sensed by UAVs when they are flying at low altitudes or placed on a ground platform. 

As a step forward to the original performance validation of the GWP and PQM algorithms, which considered a theoretical radio propagation model only, GPQM was evaluated under a realistic wireless channel model, based on experimental measurements collected in a testbed. Moreover, different traffic generation models were considered, including UDP Poisson, UDP \emph{OnOff}, and TCP \emph{Bulksend}, allowing to emulate the conditions imposed by real-world applications in multiple networking scenarios, which also goes beyond the evaluation of the original versions of PQM and GWP performed using only UDP Poisson traffic. The reference scenario, which is depicted in \cref{fig:reference-scenario}, aimed at evaluating the network performance when employing the FGW placement and the queue size derived in the illustrative example presented in \cref{sec:IllustrativeExample}. Also, GPQM was evaluated under three sets of 5 random dynamic networking scenarios composed of different numbers of FAPs -- 3 FAPs, 6 FAPs, and 12 FAPs --, aiming at considering small, medium, and large flying networks, respectively, according to the number of UAVs expected to be used in practice \cite{mozaffari2019}.

In order to evaluate the network performance when GPQM is used, two placement approaches and two state of the art traffic control algorithms were considered. The counterpart placement approaches include: 1) the FGW placed in the FAPs center (i.e., three-coordinates average considering all FAPs), which is the position that maximizes the SNR of the wireless links established between the FAPs and the FGW; and 2) the FGW placed in the center of the venue. The counterpart traffic control algorithms include: 1) RED, which drops packets based on the queue size; and 2) CoDel, which takes dropping decisions based on the time the packets are held in the queue. RED and CoDel were employing their default parameters in ns-3~\cite{AQM-ns-3:online}, but the packet size value $S$ was provided, in order to enable a fair comparison with the GPQM algorithm. The performance evaluation carried out considers two performance metrics: 
\begin{itemize}
	\item \textbf{Aggregate throughput:} the average number of bits received per second by $\text{{UAV}}_0$ (FGW).
	\item \textbf{Delay (D):} the average time taken by the packets to reach the sink application of $\text{{UAV}}_0$ since the instant of time they were generated by the source application of each $\text{FAP}_i$, which includes queuing, transmission, and propagation delays.
\end{itemize} 

Moreover, in order to assess the quality of the solutions obtained by means of the GPQM algorithm, we solved the optimization problem formulated in \cref{eq:optimization-problem} and compared the network performance achieved when employing the solutions obtained with: 1) Interior Point OPTimizer (IPOPT) \cite{wachter2006}, which is a gradient-based solver; and 2) Particle Swarm Optimization (PSO) \cite{kennedy1995}, which is a gradient-free solver. IPOPT is targeted for large-scale nonlinear constrained optimization problems defined in $n$ dimensions, which aim at minimizing an objective function $f(x)$ subject to a set of constraints. The GEKKO optimization suite \cite{beal2018gekko}, which employs IPOPT as the default solver, was used to solve the optimization problem considering this approach. In turn, PSO is inspired in the social behavior of a flock of birds. It consists of a set of $s$ candidate solutions, called particles, which are moved in an $n-$dimensional hyperspace, where $n$ is the number of parameters to be optimized. The particles are moved taking into account their position and velocity, which are updated in each iteration. The movement of each particle is influenced by its best-known local position and thrives towards the best global positions found by all particles.

\begin{figure}[!t]
	\centering
	\includegraphics[width=0.9\linewidth]{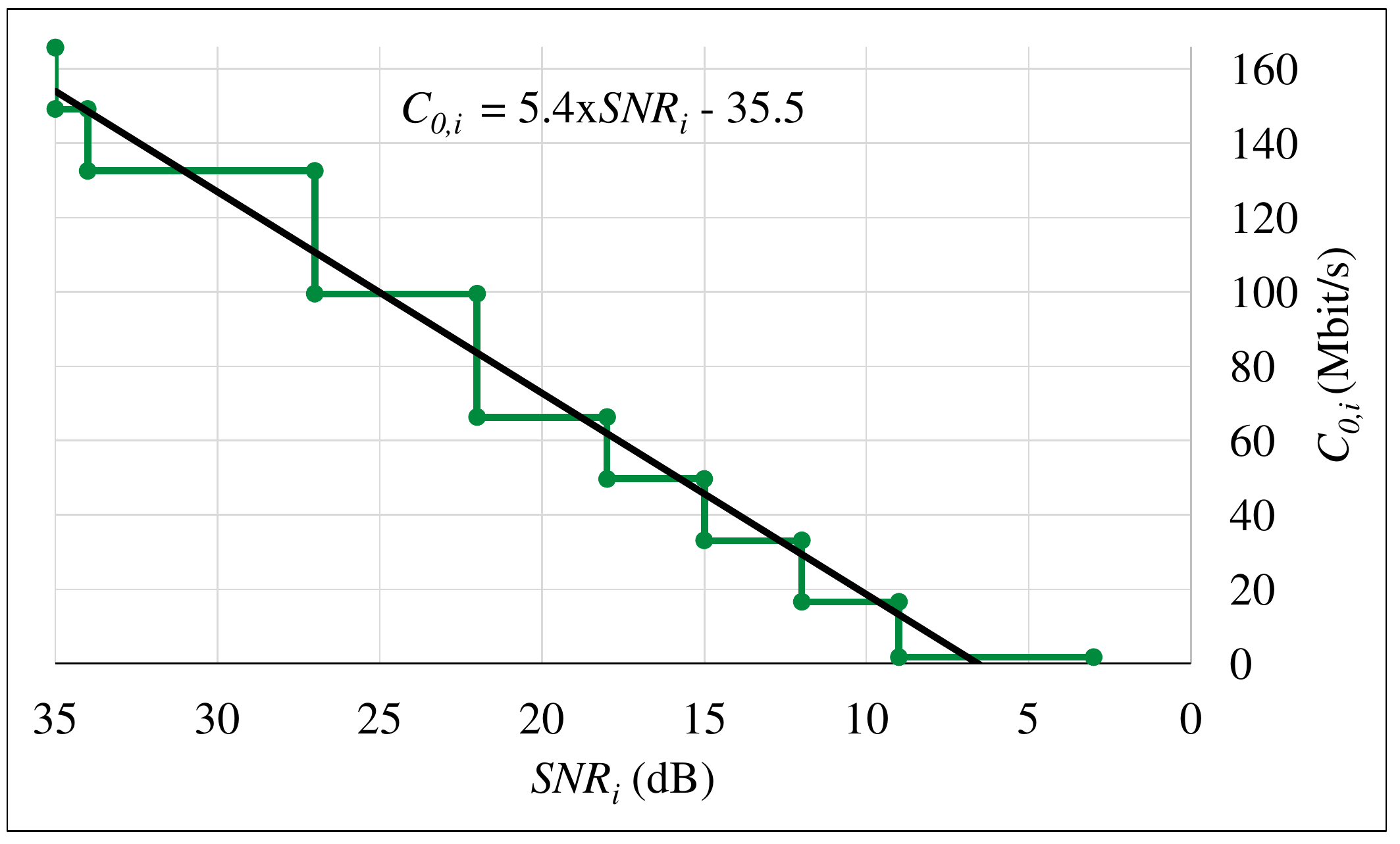}
	\caption{Wireless channel capacity modeled by a linear regression between the minimum SNR values and the data rate associated to the IEEE 802.11ac MCS indexes, considering 3 FAPs using the same wireless channel.}
	\label{fig:capacity-snr-linear-regression}
\end{figure}

For solving the optimization problem \cref{eq:optimization-problem} herein, we assumed that the capacity $C_{0,i}(t_k)$ of each wireless link results from the fair share associated to the MCS index selected by each $\text{FAP}_i$ (cf. \cref{tab:mapping-snr-channel-capacity}), which requires a minimum SNR value, following the rationale of GPQM. We modeled the relation between SNR and fair share as a continuous function using a linear regression, which is a function that closely fits the data; it is depicted in \cref{fig:capacity-snr-linear-regression} for 3 FAPs that use the same wireless channel. In order to calculate the theoretical SNR values, we employed the Friis propagation loss model (cf. \cref{eq:friis-propagation-model}), considering constant noise power, whereas the queue size and the average packet delay were determined by means of the M/D/1 queuing model (cf. \cref{eq:queue-size} and \cref{eq:waiting-time}) assuming the data packets have constant size and mutually independent inter-arrival times. This enables a fair comparison with the GPQM algorithm. Solutions were determined considering random networking scenarios composed of 3, 6, and 12 static FAPs; static FAPs were considered due to the challenges imposed by dynamic networking scenarios to the gradient-based solvers, including non-differentiable functions and constraints, multiple local minima, and the need to specify the initial values and set the variable bounds.

\subsection{Simulation Setup}
In order to evaluate the performance of GPQM in ns-3, the FAPs and the FGW were configured to use a Network Interface Card in Ad Hoc mode, considering the IEEE 802.11ac technology in channel 50, which enables \SI{160}{\mega\hertz} channel bandwidth and \SI{800}{\nano\second} Guard Interval. The wireless link between each $\text{FAP}_i$ and the FGW was employing a single spatial stream, since, due to the strong Line of Sight, the spatial diversity provided by the Multiple Input Multiple Output (MIMO) technique is not a relevant advantage for the wireless links established between UAVs. Three different traffic types were considered: 1) UDP Poisson, which is a traffic model widely used in queuing theory, since the aggregation of multiple independent flows constitutes a Poisson arrival process; 2) UDP \emph{OnOff}, in order to represent traffic sources that include active and idle periods, such as Web servers; and 3) TCP \emph{BulkSend}, which is made up of multiple bursts of consecutive packets, typically associated to a File Transfer Protocol session \cite{bidgoli2007}. %The use of different traffic types allowed to evaluate the performance of GPQM under the conditions imposed by real-world communications services and online applications. 
The data rate was defined by the \emph{IdealWifiManager} auto rate mechanism~\cite{ideal-wifi-manager:online}. The main simulation parameters used are summarized in \cref{tab:ns-3 simulation parameters}.

The FAPs were moving according to the Random Waypoint Mobility (RWM) model, with speed uniformly distributed between \SI{0.5}{\meter/\second} and \SI{3}{\meter/\second}, in order to consider time-varying capacity for the wireless links. The UAVs were placed inside a venue with dimensions $x^{MAX}=\SI{100}{\meter}$ long, $y^{MAX}=\SI{100}{\meter}$ wide, and $z^{MAX}=\SI{20}{\meter}$ high. The traffic demand of the FAPs was variable, according to the networking scenario employed, but always lower than the maximum capacity of the shared wireless channel. Since the GPQM algorithm relies on knowing in advance the positions of the FAPs, which in a real-world deployment are provided by the FAP placement algorithm running at the Central Station, we used BonnMotion \cite{aschenbruck2010}, which is a generator of mobile scenarios. The resulting waypoints, consisting of the future positions of the FAPs at each periodic time instant $t_k$, were used to calculate in advance the FGW position and the queue size for each FAP over time. Finally, the FGW position, the position and queue size for each FAP, and the transmission power for the UAVs for the whole simulation time, with a sampling period of \SI{1}{\second}, were imported to ns-3. The \emph{WaypointMobilityModel}~\cite{waypoint-mobility-manager:online} was used to place the FAPs and the FGW over time in the simulation environment, according to the positions generated by BonnMotion and defined by GPQM, respectively, which were exported to ns-3. To configure the queue size at each FAP over time, a \emph{DropTailQueue} with size scheduled in advance was implemented. 

The simulation results consider 20 runs for each scenario described in~\cref{sec:EvaluationMethodology}, under the same exact networking conditions, with \emph{RngSeed}~=~20 and \emph{RngRun}~=~\{1, ..., 20\}.

\begin{table}[!t]
	\centering
	\caption{Summary of the main ns-3 simulation parameters.}
	\label{tab:ns-3 simulation parameters}
	\begin{tabular}{lp{3cm}lp{2cm}}
		\hline
		Simulation time        & \begin{scriptsize}(\SI{30}{\second} bootstrap +)\end{scriptsize} \SI{70}{\second} 
		\\
		Wi-Fi standard         & IEEE 802.11ac                                                                     
		\\
		Wireless channel       & 50 (\SI{5250}{\mega\hertz})                                                       
		\\
		Channel bandwidth      & \SI{160}{\mega\hertz}                                                             
		\\
		Guard Interval         & \SI{800}{\nano\second}                                                            
		\\
		Propagation delay      & Constant speed                                                                    
		\\
		Propagation model      & Friis                                                                             
		\\
		Rician K-factor        & \SI{13}{dB}                                                                       
		\\
		Remote station manager & \emph{IdealWifiManager}                                                           
		\\
		Wi-Fi mode             & Ad Hoc                                                                            
		\\
		Mobility model         & Waypoint                                                                          
		\\
		Packet length          & 1400 bytes                                                                        
		\\
		Traffic types          & UDP Poisson and \emph{OnOff}, and TCP \emph{BulkSend}                             
		\\
		\hline
	\end{tabular}
\end{table}

\subsection{Simulation Results}\label{sec:simulation-results}
The obtained simulation results are the average values for the two performance metrics defined in \cref{sec:EvaluationMethodology}, computed for each second of the simulation runs. The set of average values obtained for each performance metric is represented by means of the Cumulative Distribution Function (CDF) for the packet delay and by the complementary CDF (CCDF) for the aggregate throughput. The CDF $F(x)$ represents the percentage of samples for which the delay was lower than or equal to $x$, while the CCDF $F'(x)$ represents the percentage of samples for which the aggregate throughput was higher than $x$. The analysis presented herein takes into account the results for the 90\textsuperscript{th} percentile. 

\subsubsection{Reference Scenario}\label{sec:simulation-results-static-scenario}
\begin{figure}[!t]
	\centering
	\subfloat[Aggregate throughput CCDF.]{
		\includegraphics[width=0.85\linewidth]{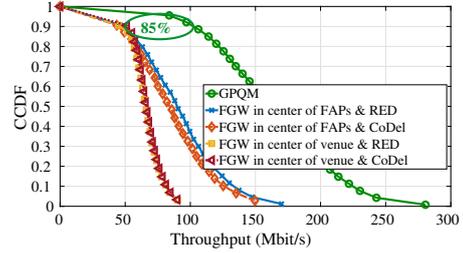}
		\label{fig:gpqm-static-poisson-rician-throughput}}
	\hfill
	\subfloat[Delay CDF.]{
		\includegraphics[width=0.85\linewidth]{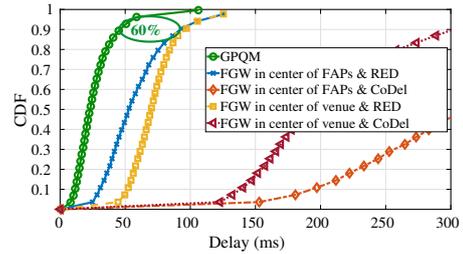}
		\label{fig:gpqm-static-poisson-rician-delay}}
	\caption{Performance results for 3 static FAPs (reference scenario of \cref{fig:reference-scenario}), considering UDP Poisson traffic.}
	\label{fig:gpqm-static-poisson-rician-results}
\end{figure}

\begin{figure}[!t]
	\centering
	\subfloat[Aggregate throughput CCDF.]{
		\includegraphics[width=0.85\linewidth]{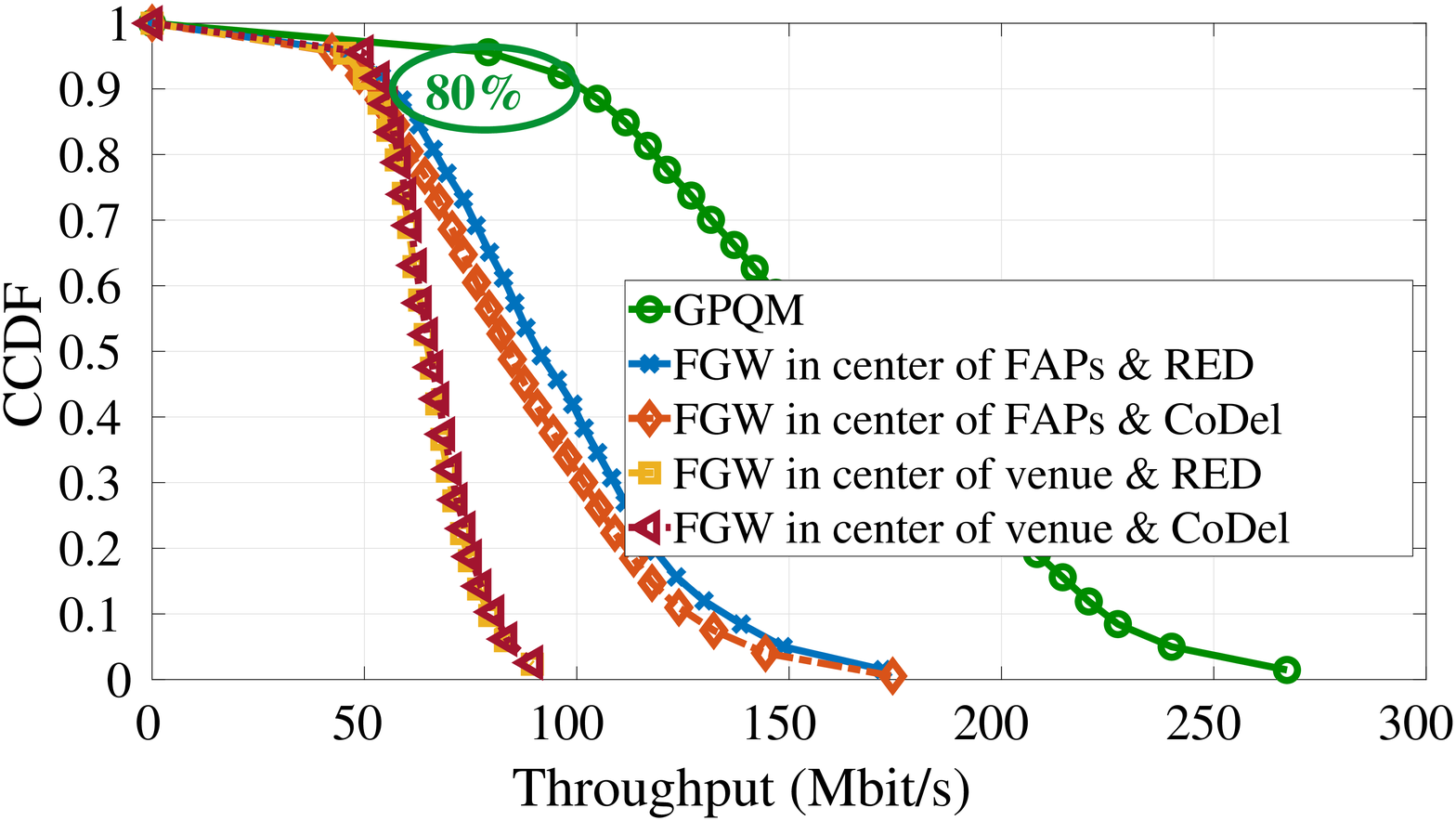}
		\label{fig:gpqm-static-onoff-rician-throughput}}
	\hfill
	\subfloat[Delay CDF.]{
		\includegraphics[width=0.85\linewidth]{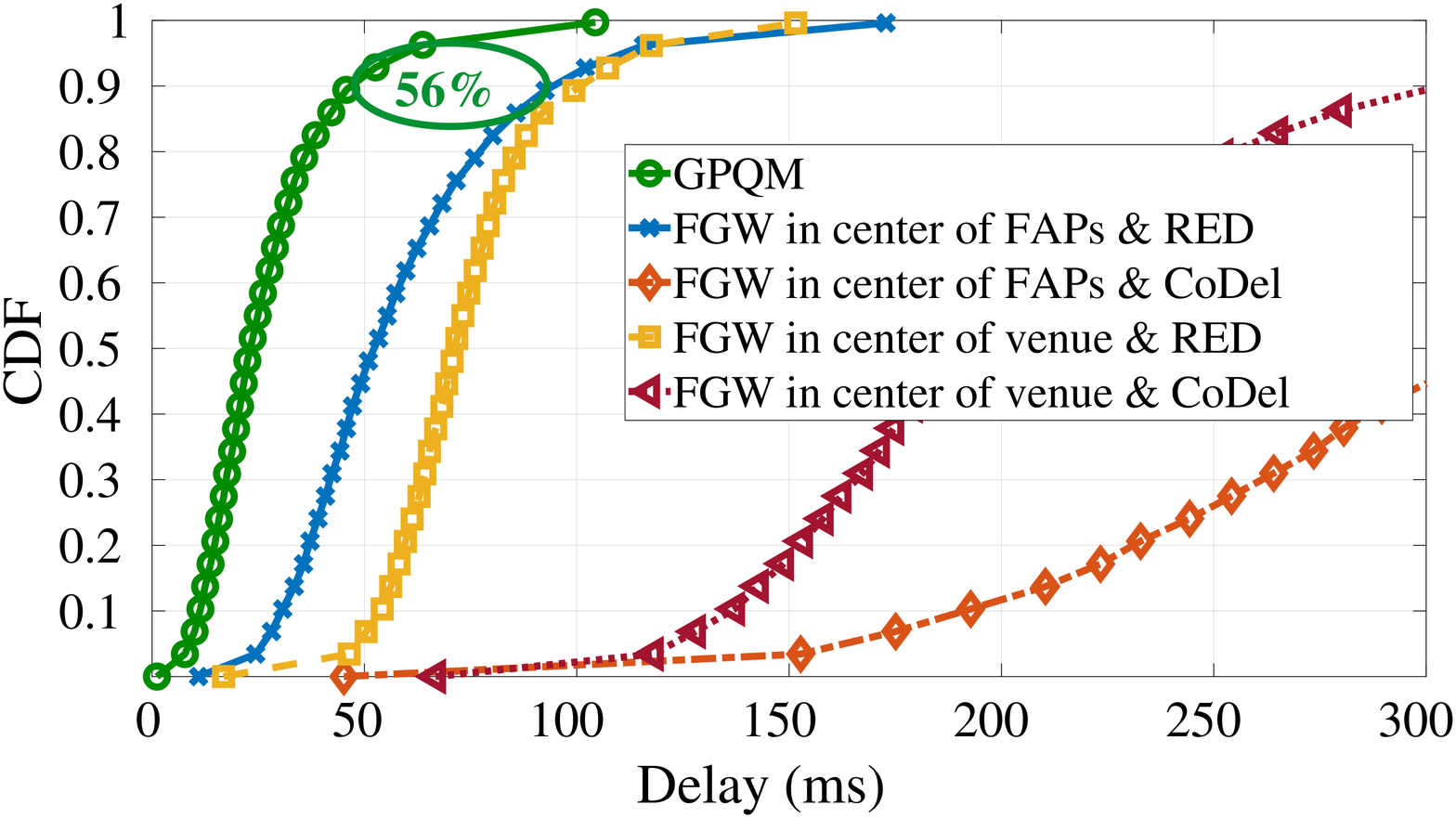}
		\label{fig:gpqm-static-onoff-rician-delay}}
	\caption{Performance results for 3 static FAPs (reference scenario of \cref{fig:reference-scenario}), considering UDP \emph{OnOff} traffic.}
	\label{fig:gpqm-static-onoff-rician-results}
\end{figure}

\begin{figure}[!t]
	\centering
	\subfloat[Aggregate throughput CCDF.]{
		\includegraphics[width=0.85\linewidth]{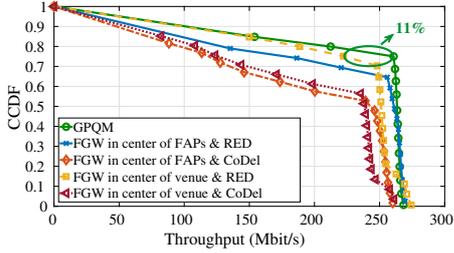}
		\label{fig:gpqm-static-bulksend-rician-throughput}}
	\hfill
	\subfloat[Delay CDF.]{
		\includegraphics[width=0.85\linewidth]{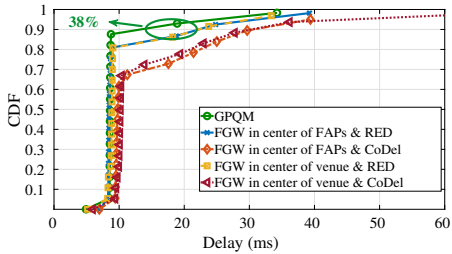}
		\label{fig:gpqm-static-bulksend-rician-delay}}
	\caption{Performance results for 3 static FAPs (reference scenario of \cref{fig:reference-scenario}), considering TCP \emph{BulkSend} traffic.}
	\label{fig:gpqm-static-bulksend-rician-results}
\end{figure}

Regarding the reference scenario (cf. \cref{fig:reference-scenario}), we conclude that GPQM allows to improve the aggregate throughput up to 85\%, considering UDP Poisson traffic (cf.~\cref{fig:gpqm-static-poisson-rician-results}), while the average delay is decreased up to 60\%. Similarly, for UDP \emph{OnOff} traffic (cf.~\cref{fig:gpqm-static-onoff-rician-results}), the average delay is decreased up to 56\%, while the aggregate throughput is improved up to 80\%. When considering TCP \emph{BulkSend} traffic (cf.~\cref{fig:gpqm-static-bulksend-rician-results}), the average delay is decreased up to 38\%, while the aggregate throughput is improved in 11\%.
GPQM is the solution that provides the best network performance. This is achieved by ensuring the smallest possible queue size for each $\text{FAP}_i$, avoiding packets from being held in the transmission queue for a long time, which significantly increases the delay of all packets without relevant gains in the aggregate throughput.

%%% GPQM | Static scenario | Poisson | Rician %%%

%%% GPQM | Static scenario | OnOff | Rician %%%

%%% GPQM | Static scenario | BulkSend | Rician %%%

\subsubsection{Dynamic Scenarios}\label{sec:simulation-results-dynamic-scenarios}
\begin{figure}[!t]
	\centering
	\subfloat[Aggregate throughput CCDF.]{
		\includegraphics[width=0.85\linewidth]{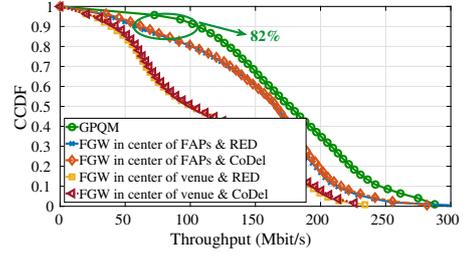}
		\label{fig:GPQM-dynamic-onoff-rician-throughput-3FAPs}}
	\hfill
	\subfloat[Delay CDF.]{
		\includegraphics[width=0.85\linewidth]{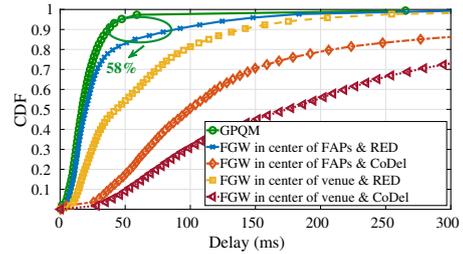}
		\label{fig:GPQM-dynamic-onoff-rician-delay-3FAPs}}
	\caption{Performance results for 3 moving FAPs, under 5 random dynamic networking scenarios, considering UDP \emph{OnOff} traffic.}
	\label{fig:GPQM-dynamic-onoff-rician-results-3FAPs}
\end{figure}

\begin{figure}[!t]
	\centering
	\subfloat[Aggregate throughput CCDF.]{
		\includegraphics[width=0.85\linewidth]{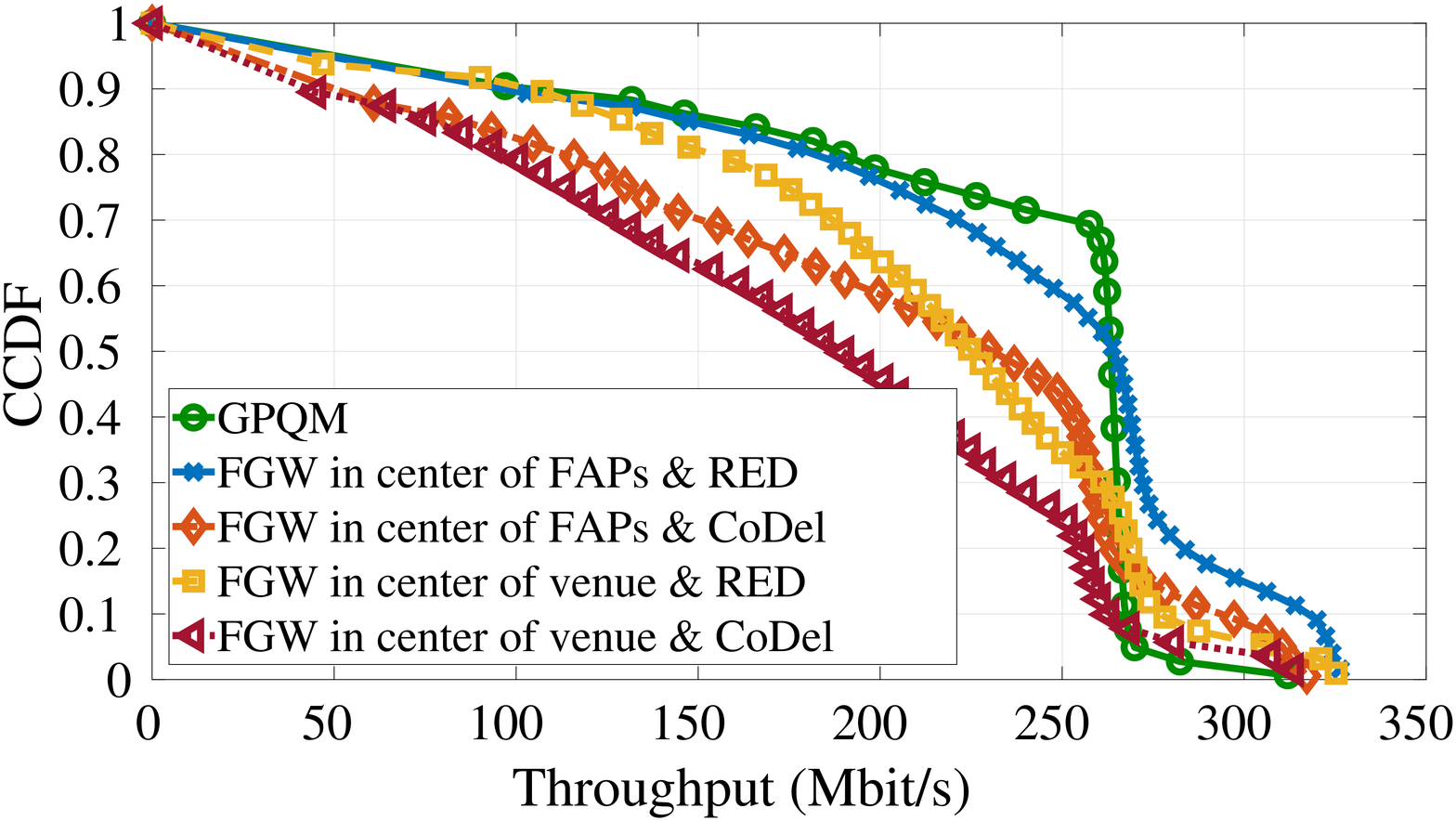}
		\label{fig:GPQM-dynamic-bulksend-rician-throughput-3FAPs}}
	\hfill
	\subfloat[Delay CDF.]{
		\includegraphics[width=0.85\linewidth]{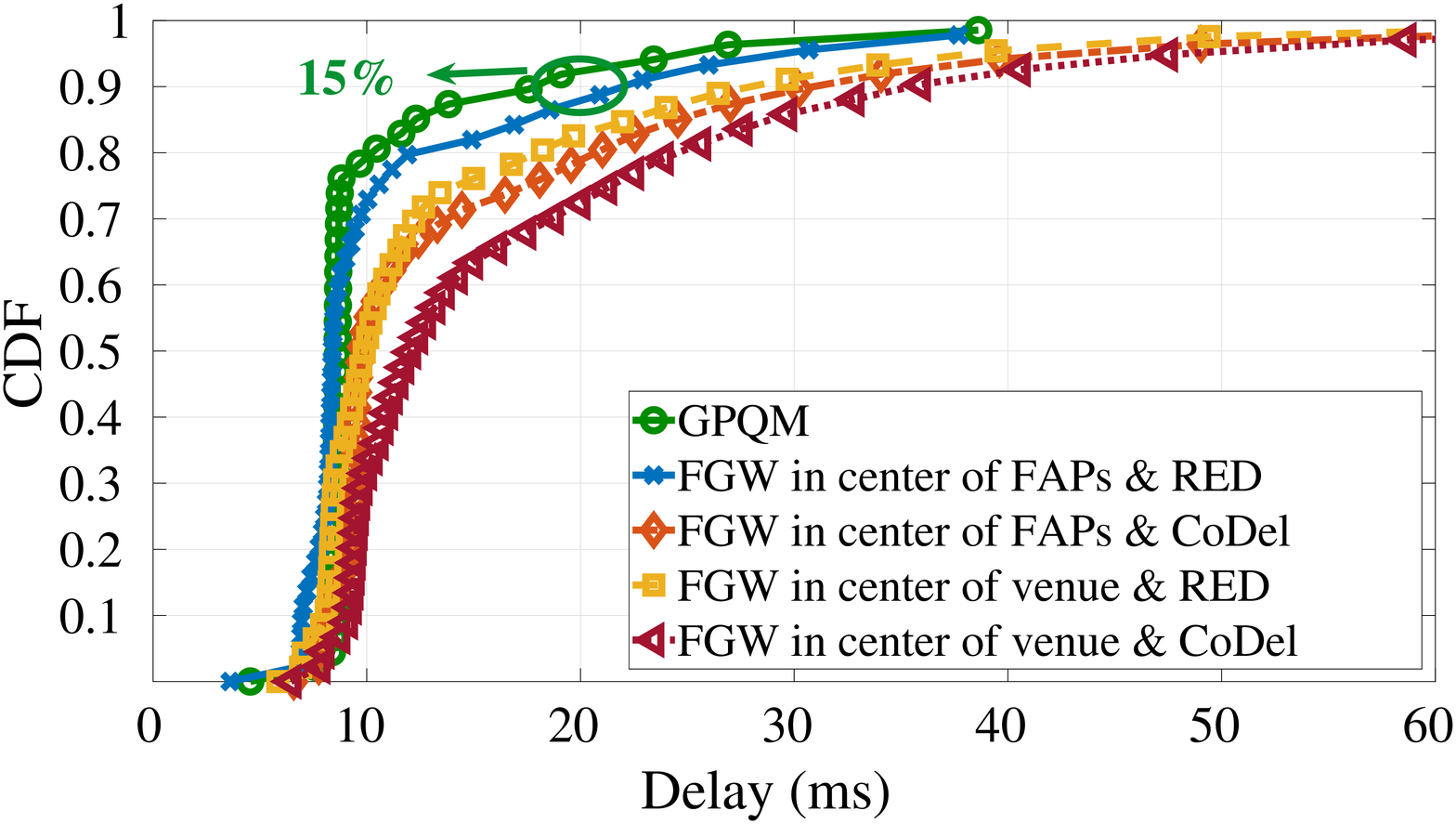}
		\label{fig:GPQM-dynamic-bulksend-rician-delay-3FAPs}}
	\caption{Performance results for 3 moving FAPs, under 5 random networking scenarios, considering TCP \emph{BulkSend} traffic.}
	\label{fig:GPQM-dynamic-bulksend-rician-results-3FAPs}
\end{figure}

\begin{figure}[!t]
	\centering
	\subfloat[Aggregate throughput CCDF.]{
		\includegraphics[width=0.85\linewidth]{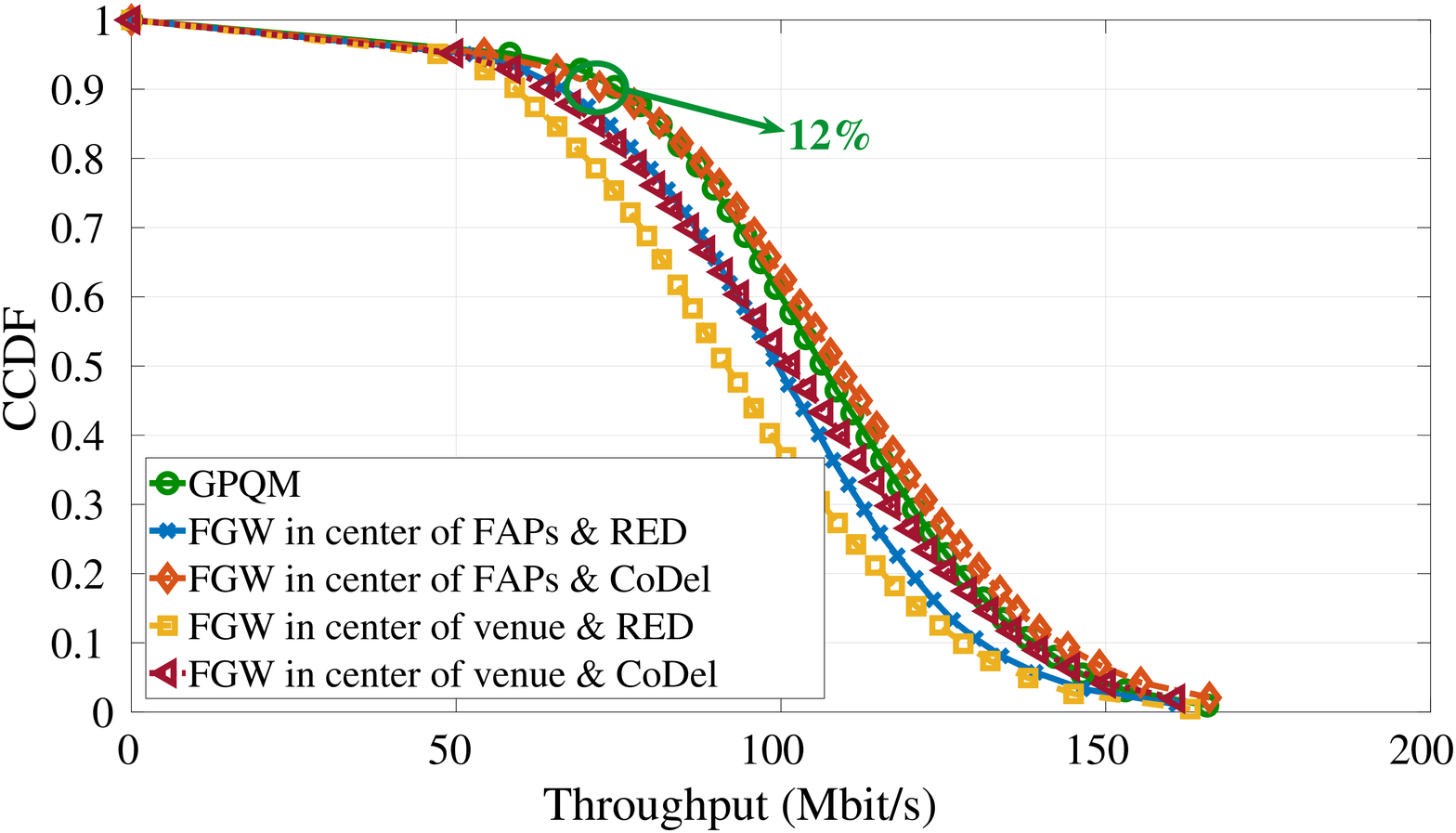}
		\label{fig:GPQM-dynamic-poisson-rician-throughput-6FAPs}}
	\hfill
	\subfloat[Delay CDF.]{
		\includegraphics[width=0.85\linewidth]{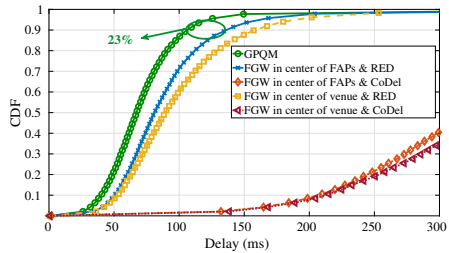}
		\label{fig:GPQM-dynamic-poisson-rician-delay-6FAPs}}
	\caption{Performance results for 6 moving FAPs, under 5 random networking scenarios, considering UDP Poisson traffic.}
	\label{fig:GPQM-dynamic-poisson-rician-results-6FAPs}
\end{figure}

\begin{figure}[!t]
	\centering
	\subfloat[Aggregate throughput CCDF.]{
		\includegraphics[width=0.85\linewidth]{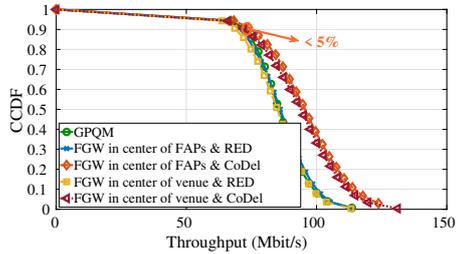}
		\label{fig:GPQM-dynamic-onoff-rician-throughput-12FAPs}}
	\hfill
	\includegraphics[width=0.85\linewidth]{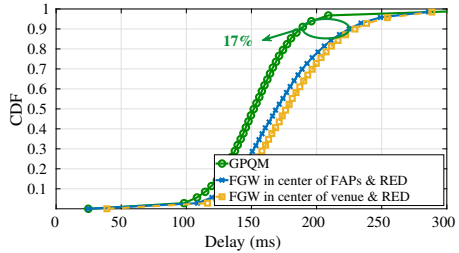}
	\subfloat[Delay CDF.]{
		\includegraphics[width=0.85\linewidth]{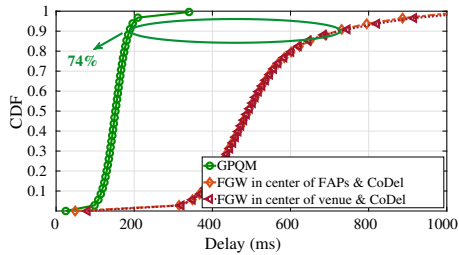}
		\label{fig:GPQM-dynamic-onoff-rician-delay-12FAPs}}
	\caption{Performance results for 12 moving FAPs, under 5 random networking scenarios, considering UDP \emph{OnOff} traffic.}
	\label{fig:GPQM-dynamic-onoff-rician-results-12FAPs}
\end{figure}

\begin{figure}[!t]
	\centering
	\subfloat[Aggregate throughput CCDF.]{
		\includegraphics[width=0.85\linewidth]{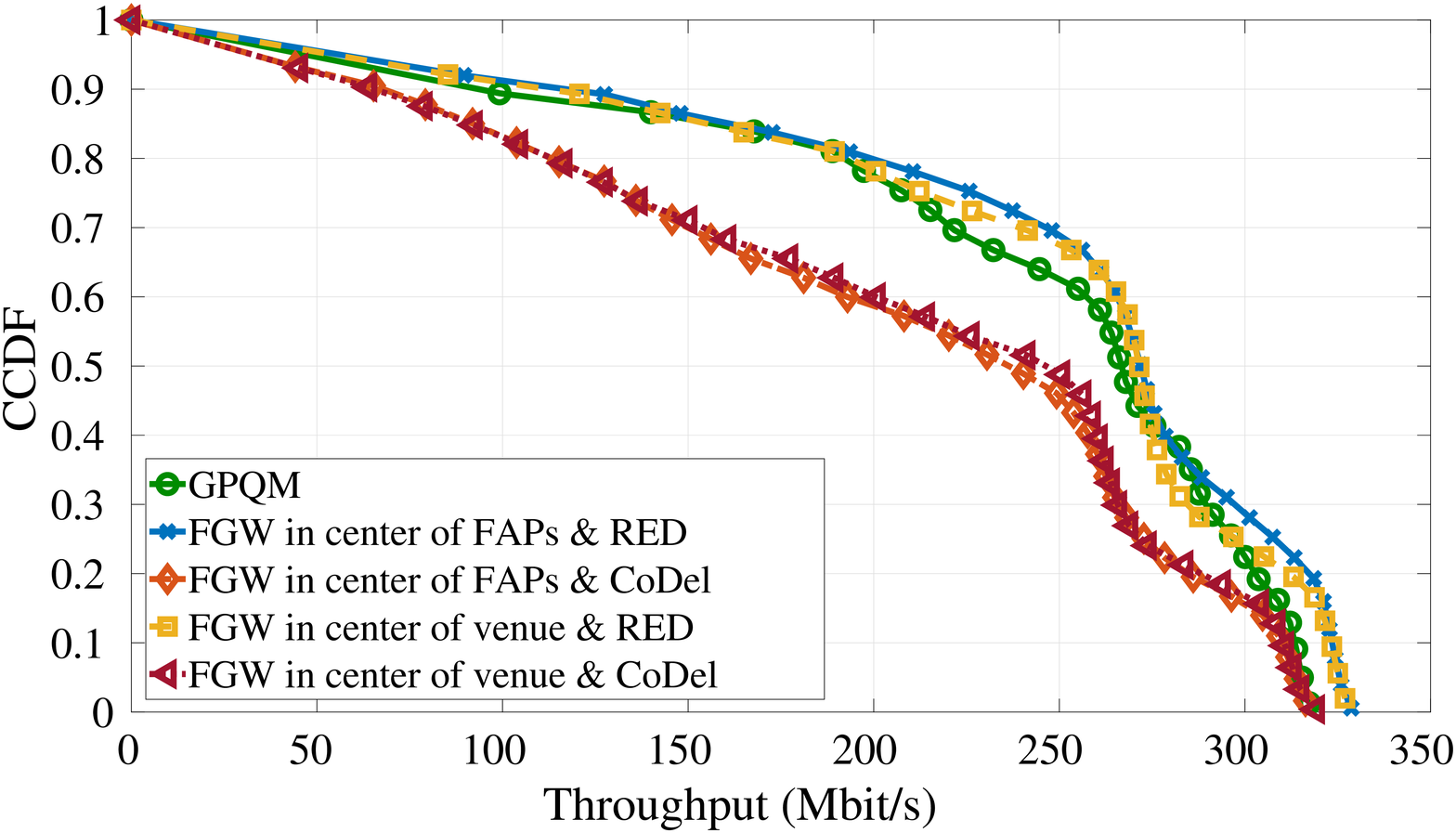}
		\label{fig:GPQM-dynamic-bulksend-rician-throughput-12FAPs}}
	\hfill
	\subfloat[Delay CDF.]{
		\includegraphics[width=0.85\linewidth]{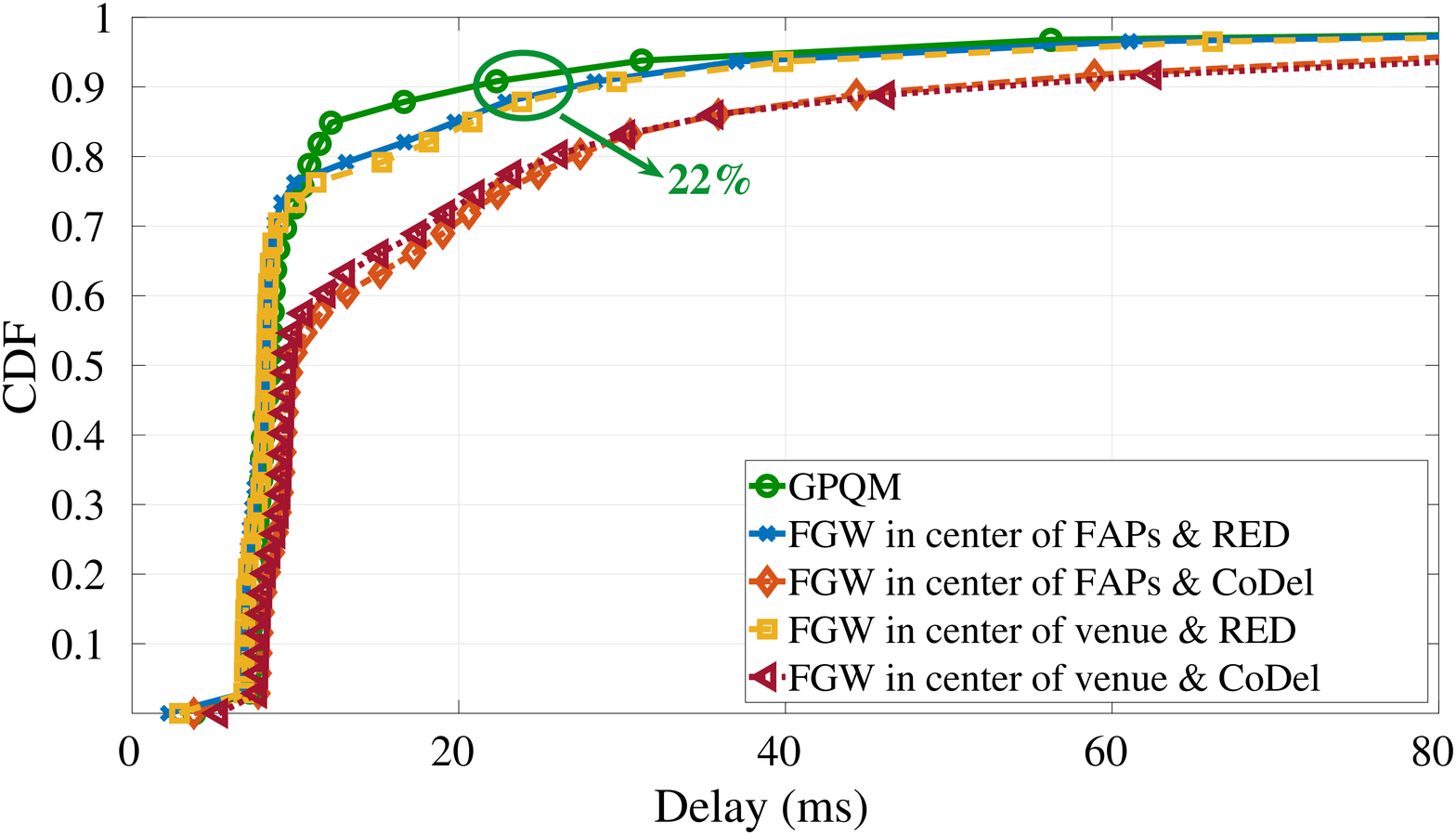}
		\label{fig:GPQM-dynamic-bulksend-rician-delay-12FAPs}}
	\caption{Performance results for 12 moving FAPs, under 5 random networking scenarios, considering TCP \emph{Bulksend} traffic.}
	\label{fig:GPQM-dynamic-bulksend-rician-results-12FAPs}
\end{figure}

The results for the dynamic scenarios, which consisted of 3, 6, and 12 FAPs in movement during \SI{70}{\second} simulation time, are depicted in 
\cref{fig:GPQM-dynamic-onoff-rician-results-3FAPs}, \cref{fig:GPQM-dynamic-bulksend-rician-results-3FAPs}, \cref{fig:GPQM-dynamic-poisson-rician-results-6FAPs}, \cref{fig:GPQM-dynamic-onoff-rician-results-12FAPs}, and \cref{fig:GPQM-dynamic-bulksend-rician-results-12FAPs}, considering UDP Poisson, UDP \emph{OnOff}, and TCP \emph{BulkSend} traffic types. GPQM improves the aggregate throughput up to 82\% while reducing the average delay up to 58\% (cf. \cref{fig:GPQM-dynamic-onoff-rician-results-3FAPs}). In the worst-case scenario, GPQM enables approximately the same aggregate throughput for a decrease in the average delay up to 15\% (cf. \cref{fig:GPQM-dynamic-bulksend-rician-delay-3FAPs}), which was observed for TCP \emph{Bulksend} traffic, considering 3 FAPs. The gains in the average delay for TCP are higher when the number of FAPs increases, as expressed by the gain of 22\% in \cref{fig:GPQM-dynamic-bulksend-rician-delay-12FAPs}. The lower gains for TCP traffic with respect to UDP traffic are justified by the fact that TCP already includes a congestion control mechanism. The gains are even more relevant if we consider CoDel as the counterpart queue management algorithm. Overall, CoDel allows to achieve high throughput, at the expense of increased average delay. Taking into account the results for UDP \emph{OnOff} traffic and 12 FAPs, the aggregate throughput achieved by GPQM is approximately 5\% lower with respect to CoDel; however, GPQM allows to decrease the average delay up to 74\% (cf. \cref{fig:GPQM-dynamic-onoff-rician-results-12FAPs}). 

The performance evaluation carried out allows to conclude that GPQM, in the worst-case scenario, enables approximately the same aggregate throughput for a lower average delay. In addition, we can infer that the influence of GPQM on performance improvement is lower when the number of FAPs increases. Yet, the number of FAPs that compose real-world flying networks is typically limited, motivating the use of GPQM in practice.\looseness=-1

%%% GPQM | Dynamic scenario | OnOff | Rician | 3 FAPs %%%

%%% GPQM | Dynamic scenario | Bulksend | Rician | 3 FAPs %%%

%%% GPQM | Dynamic scenario | Poisson | Rician | 6 FAPs %%%

%%% GPQM | Dynamic scenario | OnOff | Rician | 12 FAPs %%%

%%% GPQM | Dynamic scenario | TCP Bulksend | Rician | 12 FAPs %%%

%%% GPQM Benchmarking | Dynamic scenario (5 topologies) | Poisson | Rician %%%
\begin{figure}[!t]
	\centering
	\subfloat[Aggregate throughput CCDF.]{
		\includegraphics[width=0.90\linewidth]{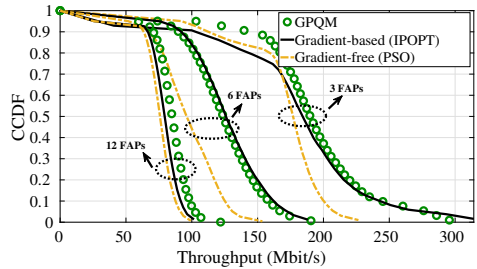}
		\label{fig:optimization-dynamic-poisson-rician-throughput}}
	\hfill
	\subfloat[Delay CDF.]{
		\includegraphics[width=0.90\linewidth]{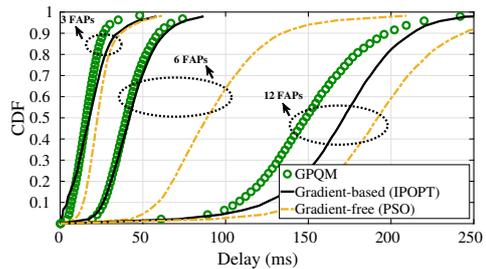}
		\label{fig:optimization-dynamic-poisson-rician-delay}}
	\caption{Benchmarking results considering the FGW position, $P_0 = (x_0, y_0, z_0)$, the transmission power of the UAVs, $P_T$, and the queue size for each $\text{FAP}_i$, $Q_i$, determined by means of 1) the GPQM algorithm (green circles), 2) the gradient-based solver (solid black lines), and 3) the gradient-free solver (dashed yellow lines), considering the optimization problem formulated in \cref{eq:optimization-problem}, for 3, 6, and 12 FAPs generating UDP Poisson traffic, under 5 random networking scenarios for each number of FAPs.}
	\label{fig:optimization-dynamic-poisson-rician-results}
\end{figure}

\subsection{GPQM Benchmarking}\label{sec:solver}
Taking into account the FGW position, $P_0 = (x_0, y_0, z_0)$, the transmission power of the UAVs, $P_T$, and the queue size for each $\text{FAP}_i$, $Q_i$, determined by means of 1) the GPQM algorithm, 2) the gradient-based solver IPOPT, and 3) the gradient-free solver PSO, ns-3 simulations were carried out to evaluate the flying network performance and assess the quality of the solutions obtained. ns-3 allows to simulate the stochastic characteristics of wireless networks and the behavior of representative traffic generation models, which are too complex to be modeled mathematically. For that purpose, it takes advantage of simulation models that represent abstractions of communications nodes, allowing to accurately evaluate the performance of wireless networks by means of a packet-driven approach. Since we observed in practice that, as the number of UAVs increases the time required to solve the optimization problem defined in~\cref{eq:optimization-problem} grows exponentially, which makes it difficult to obtain solutions for dynamic flying networks composed of UAVs in movement, static networking scenarios, in which 3, 6, and 12 FAPs were placed in fixed positions, were considered for benchmarking purposes. It is important to note that this limitation, which led to the use of static networking scenarios in this evaluation, was imposed by IPOPT and PSO; the GPQM algorithm was evaluated considering dynamic networking scenarios in~\cref{sec:simulation-results-dynamic-scenarios}. The benchmarking results are depicted in \cref{fig:optimization-dynamic-poisson-rician-results}. The plots clearly show that the GPQM algorithm (green circles) leads to the best network performance. While the solutions obtained using a gradient-free solver (PSO) provide the worst network performance (dashed yellow lines), since they are only acceptable solutions, the solutions obtained using a gradient-based solver (IPOPT) (solid black lines) are closer to the results obtained using GPQM. Yet, solving an optimization problem using a gradient-based solver implies specifying initial values, setting variable bounds, and scaling variables and equations. For different initial values, different solutions are achieved. However, the convergence to the optimal solutions relies on good directional information (i.e., good derivatives), for which providing an initial point close to a local optimum and compliant with all constraints of the optimization problem may improve the solver performance. Still, setting the initial values is not straightforward, since there is no information about the regions where the solutions of the optimization problem are located. The challenge when using gradient-based solvers is exacerbated in networks with dynamic topology, for which the initial values must be defined for each topology reconfiguration. 
GPQM employs a heuristic approach based on the theoretical domain knowledge that improving the transmission power of the communications nodes leads to higher SNR for the wireless links and improved network capacity. This is the main differentiating factor between GPQM and state of the art solvers, since the latter change the values of the variables to be optimized without domain knowledge about their influence on network performance.

%%%%%%%%%%%%%%%%%%%%%%%%%%%%%%%%%%%%%%%%%%%%%
% DISCUSSION
%%%%%%%%%%%%%%%%%%%%%%%%%%%%%%%%%%%%%%%%%%%%%
\section{Discussion \label{sec:Discussion}}
The performance evaluation carried out allows to conclude significant gains in the QoS offered by the flying network when the GPQM algorithm is used. GPQM is especially suitable for highly dynamic flying networks, which  motivate adjustments in the FGW placement over time, in order to meet the traffic demand of the FAPs, as well queues with dynamic size, for addressing the changes in the packet service time that are induced by the time-varying capacity of the wireless links. The use of GPQM is especially useful in networks at the onset of congestion, typical in crowded and emergency scenarios. The GPQM algorithm was designed for networking scenarios composed of a single FGW; however, it can also be employed in networking scenarios with multiple FGWs, which should be configured to use orthogonal wireless channels, in order to minimize interference. This is especially relevant when 1) a single wireless channel does not provide high enough capacity to meet the traffic demand of all FAPs in the network and 2) the FAPs are placed over a wide area, such that a FAP may get out of reach of the FGW it is currently assigned to. For that purpose, clusters of closer FAPs can be formed, and for each cluster of FAPs a dedicated FGW may be assigned. In this sense, the GPQM algorithm can be employed to define the position of the FGW of each cluster of FAPs, as well as the FAPs' queue size. The overall network performance will be improved if the multiple FGWs are able to use different connections to the Internet (e.g., cellular Base Stations and tethered drones).

GPQM provides the same output whether downlink or uplink traffic is considered, since all the UAVs are configured with the same transmission power and the wireless channel is assumed to be symmetric. In this sense, for the downlink traffic, the FGW may include a queue for each FAP and take advantage of a scheduling mechanism to select which queue to remove a packet from over time, in order to ensure fairness. This paves the way to use the GPQM algorithm in emerging networking scenarios where symmetric traffic applications are growing~\cite{elshaer2014}, such as social networks, video streaming, and online gaming. In networking scenarios with asymmetric wireless links, the lowest capacity among the two directions should be considered for placing the FGW. 

The position of the FGW and the queue size of the FAPs are computed by the GPQM algorithm for discrete time instants. Each time instant corresponds to a snapshot of the flying network where the FGW and the FAPs are hovering in fixed positions (i.e., zero flying speed), in order to serve the ground users. The periodic time instants considered by the GPQM algorithm are defined based on the update period of the state of the art FAP placement algorithm used. Considering discrete time instants for computing up-to-date decisions for both the UAV placement and queue management allows to consider a trade-off between: 1) the stability of the flying network; 2) the time required by the algorithms for computing the decisions; 3) the time taken by the UAVs to reconfigure themselves according to the up-to-date decisions; and 4) the network performance gains obtained in practice. If the update period is set to a high value, the flying network will be more stable but will take a longer time to react to changes in the QoS levels demanded by the ground users. Conversely, if the update period is set to a low value, the flying network will react quicker to changes in the QoS levels demanded by the ground users. However, the flying network will be less stable, possibly making it impossible for UAVs to position themselves as fast as the decisions computed by the algorithms, which may result in negligible or no gain in practice.

It must be noted that, since the transmission power $P_T$ is the fine-tuning parameter to achieve a suitable FGW position, in networking scenarios composed of 1) multiple FAPs placed over a wide area and 2) FAPs with traffic demand close to or higher than the maximum channel capacity, $P_T$ will tend to infinite, which is not feasible in a real-world deployment. In order to address this challenge, the evolution of the GPQM algorithm for networking scenarios composed of UAV relays able to forward traffic between them, thus enabling short-range wireless links, is worthy of being considered. For that purpose, the UAV relays should be configured to form multi-hop wireless paths using different communications channels, in order to mitigate the inter-flow interference and enhance the overall capacity of the flying network, while employing reduced transmission power values.

The networking scenarios employed in the evaluation of GPQM were defined so that the FAPs listen to each other, avoiding the hidden node problem. In order to avoid the hidden node problem in real-world flying networks, where the scenarios are not fully controlled, the GPQM algorithm is worthy of being improved so that the transmission power defined for the FAPs is high enough not only to ensure the minimum SNR required to meet the traffic demand of the FAPs, but also to ensure the FAPs listen to each other. Since the traffic-aware gateway placement and queue management problem is our focus, the Medium Access Control challenges are left for future work.

The performance evaluation of the GPQM algorithm, which was carried out under realistic channel conditions for multiple networking scenarios, allows to validate GPQM and confirm the research hypothesis claimed by this article: the performance of the flying network is improved by defining in advance 1) the FGW position and 2) the queue size for each FAP, considering both the positions of the FAPs and their traffic demand over time.

%%%%%%%%%%%%%%%%%%%%%%%%%%%%%%%%%%%%%%%%%%%%%
% CONCLUSIONS
%%%%%%%%%%%%%%%%%%%%%%%%%%%%%%%%%%%%%%%%%%%%%
\section{Conclusions \label{sec:Conclusions}}
This article proposed a traffic-aware gateway placement and queue management algorithm for flying networks, called GPQM. Taking advantage of the information provided by a state of the art FAP placement algorithm, in charge of determining the positions of the FAPs to meet the traffic demand of the ground users, GPQM proactively defines over time the FGW position and the queue size for each FAP. The performance evaluation of GPQM carried out by means of ns-3 simulations, considering a realistic wireless channel model built upon experimental results collected in a testbed, showed its ability to maximize the aggregate throughput measured in the FGW, while providing stochastic delay guarantees for the transported traffic, allowing to achieve significant gains with respect to its state of the art counterparts.

As future work, we aim at exploring the GPQM algorithm in multi-hop flying networks composed of UAV relays able to forward traffic between themselves. Besides ensuring the placement of multiple UAV relays, taking into account the traffic-aware approach followed by the GPQM algorithm, we aim at combining the proposed queue management approach with the RedeFINE routing protocol that we have proposed in \cite{Coelho2018}, in order to enable multi-hop paths composed of communications nodes configured with a suitable queue size.

%%%%%%%%%%%%%%%%%%%%%%%%%%%%%%%%%%%%%%%%%%%%%%%%%%%%%%%%%%%%%%%%%
% APPENDIX
%%%%%%%%%%%%%%%%%%%%%%%%%%%%%%%%%%%%%%%%%%%%%%%%%%%%%%%%%%%%%%%
\appendix
\section{Assessment of the Quality of the Solutions Obtained for the FGW Placement~\label{appendix:a}}
\begin{figure}[!t]
	\centering
	\subfloat[FAPs' transmission ranges (circumferences) \textbf{completely overlapping} each other.]{
		\includegraphics[width=0.70\linewidth]{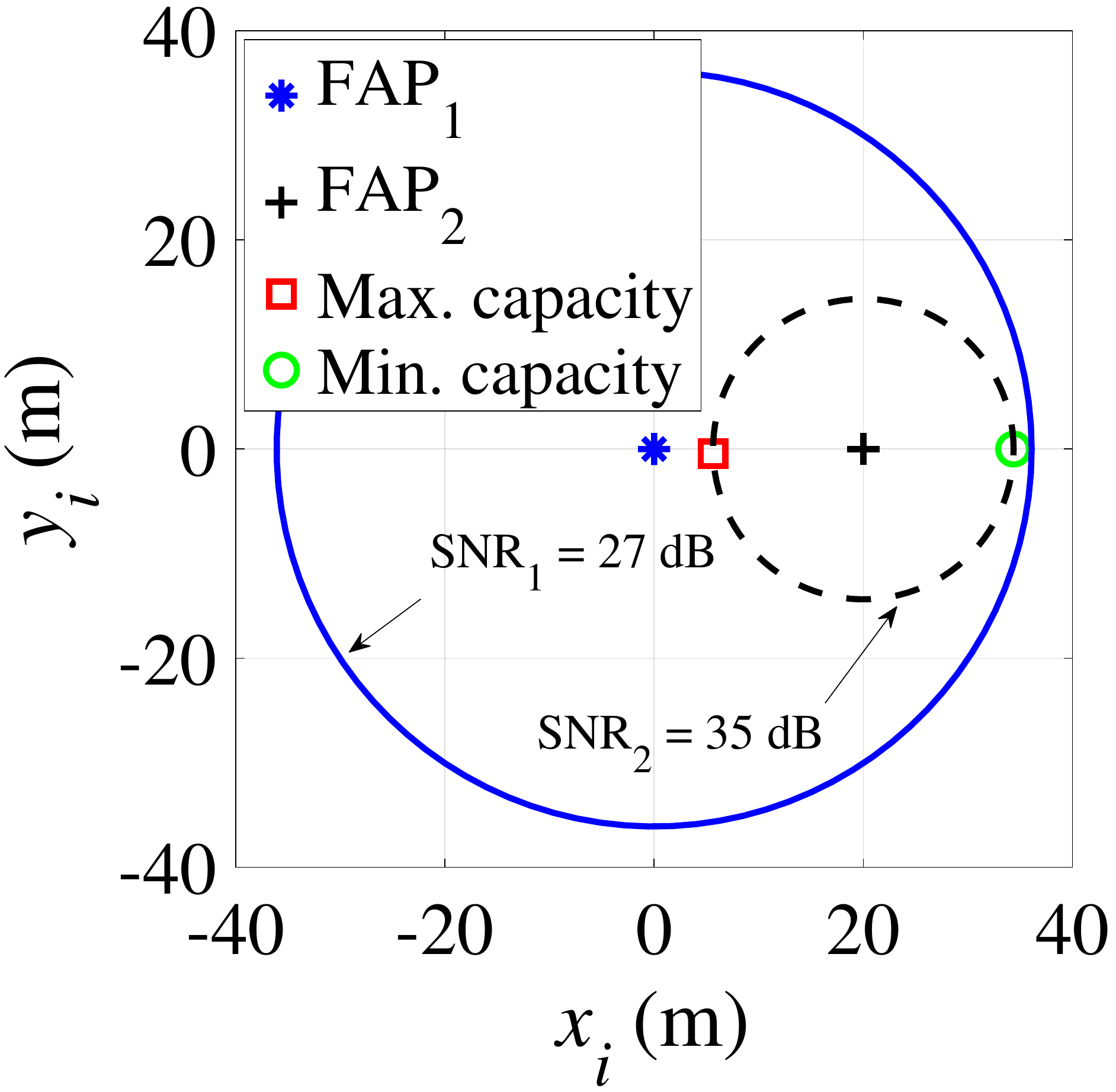}
		\label{fig:max-min-capacity-sphere-in}}
	\hfill
	\subfloat[FAPs' transmission ranges (circumferences) \textbf{partially overlapping} each other.]{
		\includegraphics[width=0.70\linewidth]{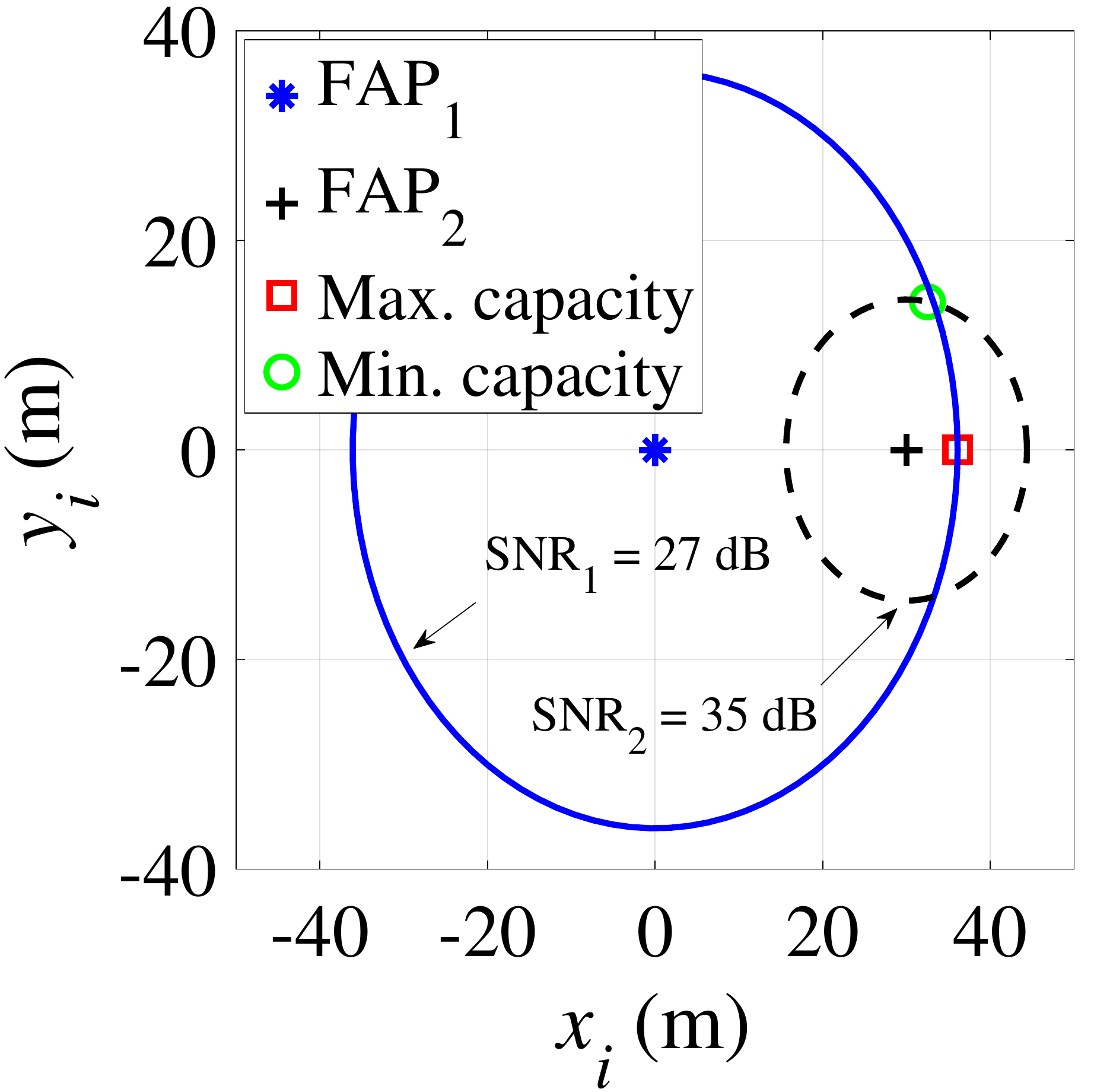}
		\label{fig:max-min-capacity-sphere-out}}
	\caption{Admissible FGW positions for maximum and minimum overall capacity $C(t_k)$ values (red square and green circle, respectively).}
	\label{fig:max-min-capacity}
\end{figure}

In order to assess the quality of the admissible solutions for the FGW placement obtained when solving the optimization problem defined in~\cref{eq:optimization-problem}, let us consider in~\cref{fig:max-min-capacity} two representative networking scenarios, composed of 2 FAPs placed at different distances from each other. They constitute the two possibilities for which at least one solution for the FGW placement can be determined: fully overlapping circumferences and partially overlapping circumferences. When the circumferences associated with the FAPs do not intersect each other, there is no solution for the problem. The circumference associated with each $\text{FAP}_i$ represents the maximum distance within which the FGW can be placed, so that the target capacity $C_{0,i}(t_k)$ demanded for the wireless link available between $\text{FAP}_i$ and the FGW is guaranteed. A larger circumference is associated to a lower SNR value, which enables a lower capacity $C_{0,i}(t_k)$, according to $B \cdot log_2 \left (1+10^\frac{SNR}{10}\right)$, given by the Shannon-Hartley theorem~\cite{Hartley1928}. For multiple FAPs, the FGW should be placed within the intersection of all the FAPs' circumferences. The SNR is determined by means of the Friis propagation loss model. In this illustrative example, we considered $B$ equal to \SI{160}{\mega\hertz}, $P_T$ set to \SI{20}{dBm}, \SI{5250}{\mega\hertz} as the carrier frequency $f$, and constant noise power $P_N$ equal to \SI{-85}{dBm}.

Admissible positions for the FGW placement are depicted by the green circles. In addition, admissible positions if the optimization problem aimed at maximizing the overall capacity $C(t_k)$ are represented by means of the red squares for comparison purposes, since maximization is a typical state of art approach in optimization problems that aim at improving the overall network performance. 

We conclude that, by minimizing the overall capacity $C(t_k)$ in our objective function, a suitable solution is obtained, since the FGW is placed in positions (green circles) that ensure the minimum targeted capacity $C_{0,i}(t_k)$ for each $\text{FAP}_i$, which is limited by the circumferences (worst-case positions). On the other hand, if the positions associated with the maximum capacity (red squares) were chosen, by maximizing the solution in our objective function, the FGW would be placed closer to one of the FAPs, but from the perspective of the remaining FAP the FGW would be placed over the corresponding circumference, which also represents a worst-case position as it occurs when minimizing the overall capacity $C(t_k)$.

For these reasons, we argue that minimizing the objective function~\cref{eq:objective-function} allows to obtain a suitable position for the FGW placement, using the minimum amount of communications resources, in contrast when the objective function is maximized. In order to take into account deviations between the theoretical and experimental values of $T_i(t_k)$ and $C_{0,i}(t_k)$ in real-world networking scenarios and accommodate bursty traffic, the value defined for $C_{0,i} (t_k)$ should include a margin with respect to the value of $T_i(t_k)$. This margin can be adjusted at each flying network reconfiguration; its fine-tuning is left for future work.

%%%%%%%%%%%%%%%%%%%%%%%%%%%%%%%%%%%%%%%%%%%%%
% ACKNOWLEDGMENTS
%%%%%%%%%%%%%%%%%%%%%%%%%%%%%%%%%%%%%%%%%%%%%
\section*{Acknowledgments}
This work is co-financed by the ERDF -- European Regional Development Fund through the Operational Programme for Competitiveness and Internationalisation -- COMPETE 2020 and the Lisboa2020 under the PORTUGAL 2020 Partnership Agreement, and through the Portuguese National Innovation Agency (ANI) as a part of the projects "CHIC: POCI-01-0247-FEDER-024498" and "5G: POCI-01-0247-FEDER-024539". The  first author also thanks the funding from FCT under the PhD grant SFRH/BD/137255/2018.

%%%%%%%%%%%%%%%%%%%%%%%%%%%%%%%%%%%%%%%%%%%%%%%%%%%%%%%%%%%%%%%%%
% REFERENCES
%%%%%%%%%%%%%%%%%%%%%%%%%%%%%%%%%%%%%%%%%%%%%%%%%%%%%%%%%%%%%%%%%
\bibliographystyle{elsarticle-num}
\bibliography{References}

\end{document}